%% file: sample-acmsmall.tex
\PassOptionsToPackage{table}{xcolor}
\documentclass[manuscript,screen]{acmart}

\usepackage[capitalize,noabbrev]{cleveref}
\usepackage{tabulary}
\usepackage{threeparttable}
\usepackage{enumitem}

\AtBeginDocument{%
  \providecommand\BibTeX{{%
    \normalfont B\kern-0.5em{\scshape i\kern-0.25em b}\kern-0.8em\TeX}}}

\setcopyright{acmcopyright}
\copyrightyear{2022}
\acmYear{2022}
\acmDOI{XXXXXXX.XXXXXXX}

\acmJournal{JACM}
\acmVolume{37}
\acmNumber{4}
\acmArticle{111}
\acmMonth{8}

\begin{document}

\title{Reflow: Automatically Improving Touch Interactions in Mobile Applications through Pixel-based Refinements}
\author{Jason Wu}
\authornote{This work was done while Jason Wu was an intern at Apple.}
\affiliation{
  \institution{HCI Institute, Carnegie Mellon University}
  \city{Pittsburgh}
  \state{PA}
  \country{USA}
}
\email{jsonwu@cmu.edu}
\author{Titus Barik, Xiaoyi Zhang, Colin Lea, Jeffrey Nichols, Jeffrey P. Bigham}
\affiliation{
  \institution{Apple}
  \city{Cupertino}
  \state{CA}
  \country{USA}
}
\email{{tbarik, xiaoyiz, colin_lea, jwnichols, jbigham}@apple.com}

\renewcommand{\shortauthors}{Wu et al.}
\begin{abstract}
Touch is the primary way that users interact with smartphones. However, building mobile user interfaces where touch interactions work well for all users is a difficult problem, because users have different abilities and preferences. We propose a system, Reflow, which automatically applies small, personalized UI adaptations, called \emph{refinements}---to mobile app screens to improve touch efficiency. Reflow uses a pixel-based strategy to work with existing applications, and improves touch efficiency while minimally disrupting the design intent of the original application. 
Our system optimizes a UI by \textit{(i)} extracting its layout from its screenshot, \textit{(ii)} refining its layout, and \textit{(iii)} re-rendering the UI to reflect these modifications. 
We conducted a user study with 10 participants and a heuristic evaluation with 6 experts and found that applications optimized by Reflow led to, on average, 9\% faster selection time with minimal layout disruption. The results demonstrate that Reflow's refinements useful UI adaptations to improve touch interactions.
\end{abstract}

\begin{teaserfigure}
    \centering
    \includegraphics[width=\textwidth]{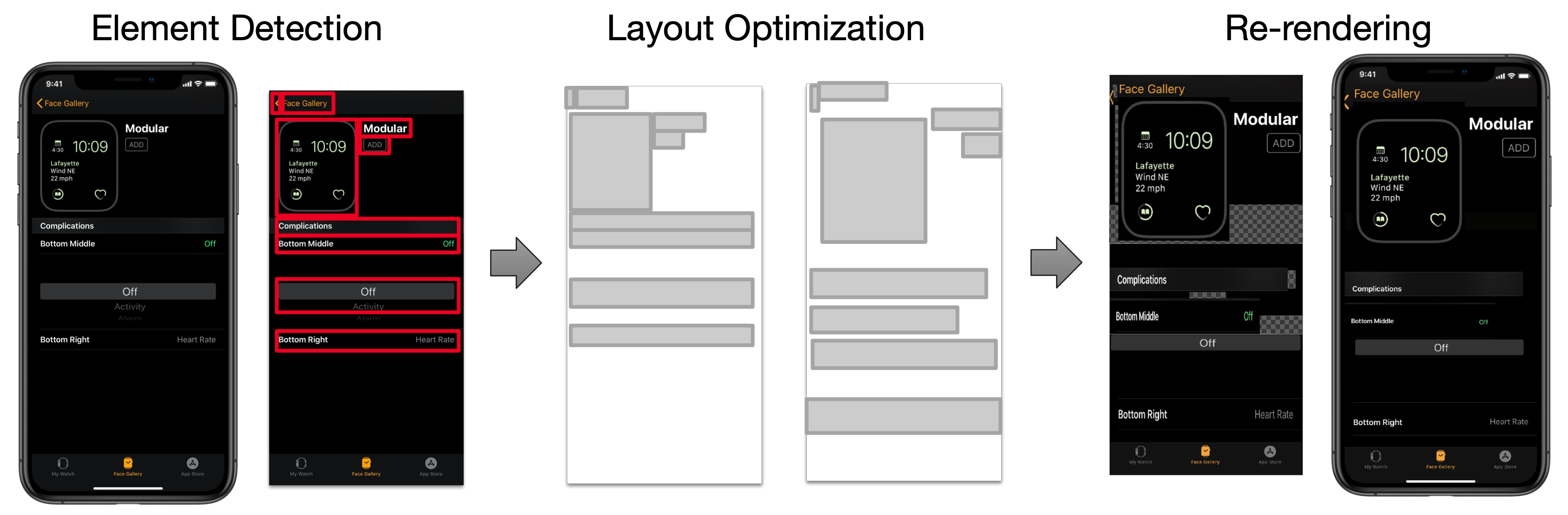}
    \captionof{figure}{Reflow {\em (i)} detects UI elements from pixels, {\em (ii)} optimizes the UI layout for a personalized difficulty model, and then {\em (iii)} re-renders the visual UI with the new layout. In this example, the UI elements are correctly detected, and the new layout includes larger buttons that are more spread apart. Simply moving and stretching the UI elements causes gaps and distortions, which Reflow fixes using additional post-processing methods.}
    \label{fig:systemoverview}
\end{teaserfigure}
\maketitle

\section{Introduction}

Touch is an ubiquitous way of interacting with smartphones. However, building mobile user interfaces (UI) where touch interactions work well for all users is a difficult problem, because users have different motor abilities, skills, and even preferences \cite{wobbrock2011ability}. For example, consider a right-handed user who wants to access a menu item on the left side of the screen. For larger screens, this menu item is more difficult for the user to touch with their right hand than for a left-handed user. 

UI adaptivity is a promising approach towards improving touch interactions, because it allows systems to dynamically personalize the UI and tailor the UI to the users’ needs. But for UI adaptivity to be practically useful for real-world apps, it must support two goals. First, the technique should be generally useful across a broad range of existing mobile applications. Second, the technique should apply adaptations in a way that respects the design intentions of the original applications. In other words, we expect that drastic UI adaptations are likely to make the user interface less familiar to the user and disruptive to the overall user experience~\cite{opportunistic}.

To operationalize these goals, we built a system, \emph{Reflow}, which automatically applies small UI adaptations---called \emph{refinements}---to mobile existing app screens to improve touch efficiency. Towards the first goal of supporting broad applicability, Reflow is entirely pixel-based: the system does not need knowledge of the applications’ dependencies or view hierarchy to make its UI adaptations. Towards the second goal of respecting design intent and minimally disrupting the user experience, Reflow incorporates the theory of microstrategies in its model~\cite{gray2000milliseconds, everett2004unintended, eng2006generating}. Microstrategies suggest that even small, principled adaptations to the user interface can significantly improve task efficiency—particularly over cumulative usage---and we postulate that the same principle applies when personalizing touch-based mobile applications.

Reflow supports personalized optimization by constructing a spatial map from usage data, which identifies difficult-to-access areas of the screen (\textit{e.g.,} elements on edges of the screen requires users to reach and reposition their hand to select).
Reflow then {\em (i)} automatically detects the UI elements contained on the screen, {\em (ii)} uses a machine learning model to optimize the UI layout to better support the difficulty map, and then {\em (iii)} re-renders the existing UI pixels to match the new layout (Figure \ref{fig:systemoverview}).
Reflow improves on existing approaches because it works with a range of existing mobile applications and enables an end-to-end pipeline from layout optimization to re-rendering the application screens.

To evaluate Reflow, we first conducted a study with 10 participants, where we found it improved interaction speed by 9\% on average, and improved interaction speeds by up to 17\% for some UIs. From lessons learned, we made further improvements to this model by detecting and applying an additional set of UI constraints (relative positioning, alignment). We then conducted a heuristic evaluation based heuristics for evaluating UI layouts \cite{todi2016sketchplore, shiripour2021grid} with 3 accessibility and 3 design experts to validate if these improvements make acceptable trade-offs between touch efficiency and layout preservation. Feedback from our expert evaluators indicated that refinements were likely to improve selection time while avoiding significant disruption to the UI. The results of this work demonstrate that refinements are a useful UI adaptation technique to improve task efficiency for touch interactions.

The contributions of our paper are as follows:
\vspace{-0.3pc}
\begin{itemize}
    \item We propose an approach based on the theory of microstrategies \cite{gray2000milliseconds} for improving touch interactions through \emph{refinements}, which are small modifications to the original UI.
    Based on this approach, we present Reflow, an end-to-end system for personalizing any existing mobile app using only its pixels.
    We describe our implementation in several modular steps: \textit{(i)} element detection, \textit{(ii)} layout refinement, \textit{(iii)} UI re-rendering, which may be applied to other UI adaptation systems.
    \item We conduct two evaluations that provide evidence for the effectiveness of Reflow's automatic UI refinements. From our user study ($n=10$), we find that the refinements automatically applied by Reflow result in more efficient touch interaction (average speedup of 9\%, up to 32\%). Furthermore, we conduct a heuristic evaluation with 3 accessibility and 3 design experts, validating that the changes induced by Reflow are likely to improve selection time while being minimally disruptive. Qualitative feedback from our expert evaluators provide additional rationale for the acceptability of these trade-offs.
\end{itemize}
\begin{figure}[t!]
\centering
\includegraphics[width=0.89\columnwidth]{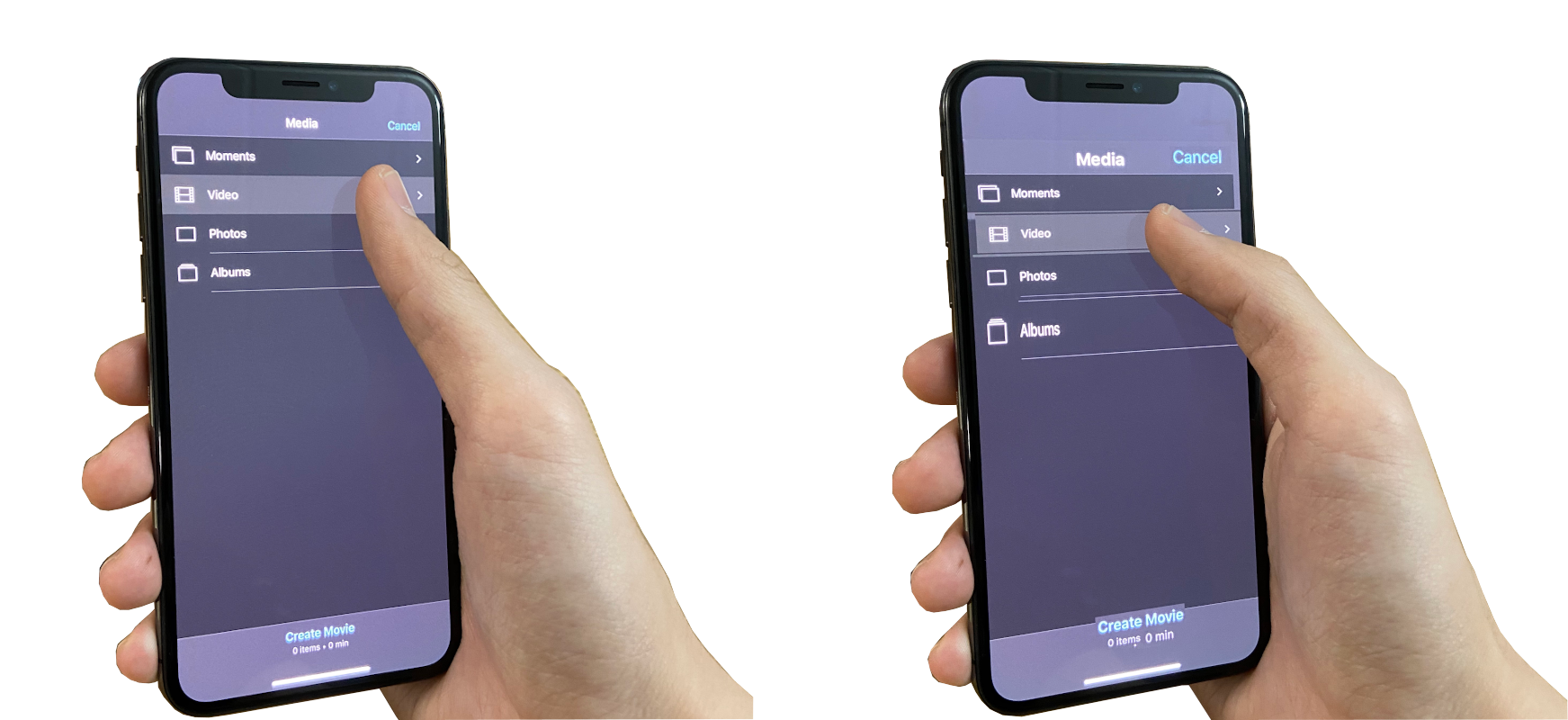}
\caption{Reflow optimizes existing third party apps using only UI information from pixels and a spatial difficulty map corresponding to a user's abilities. An app screen with short menu items near the top of the screen is difficult to use (Left). Reflow automatically optimizes the layout of on-screen elements, making each menu item taller and shifting items down (Right).}
\label{fig:reflow-hands}
\end{figure}

\section{Example Usage Scenario for Reflow}
\label{sec:example_scenario}

To motivate how Reflow can be used to improve touch interactions on mobile apps, we provide an expected usage scenario where Reflow would be enabled globally through the mobile operating system.

\textbf{Scenario:}
Alice has recently purchased a new smartphone. Alice's new smartphone is able to install and run all of her favorite apps, but the device is slightly larger than her old smartphone. Because of this, she can no longer comfortably reach UI controls on the far edges of the screen while using the device with one hand. While she is still able to use her favorite apps, Alice finds her routine interactions with the apps inconvenient.

The Reachability accessibility feature built into iOS\footnote{https://support.apple.com/guide/iphone/touch-iph77bcdd132/14.0/ios/14.0\#iph145eba8e9} addresses the challenge of touching items in the upper half of the screen by enabling users to swipe down on the lower portion of the screen to move the upper half of the screen down to the lower half. While this is useful to many people, it requires an additional touch interaction (swipe down) which slows Alice down and feels like more support than she requires.

\textbf{Setup:}
Alice opens the settings application on her smartphone and turns on the ``Reflow" toggle button. If this is the first time the setting has been enabled, Alice is brought to a setup screen that that initializes the Reflow system. This screen is similar to the first-time setup process of some biometric authentication features (\textit{e.g.,} Face ID). The setup screen asks Alice to perform a calibration task, and measures how quickly she can select targets located at different parts of the screen. With this usage information, Reflow creates a personalized profile for her that identifies difficult-to-reach areas on the screen.

\textbf{Usage:}
Once enabled, Reflow automatically intercepts and applies refinements to app screens before they are displayed to the user, a form of \emph{manifest interface} \cite{prefab}. The UI produced by Reflow is interactive, as it automatically redirects input events to the original app screen, using techniques similar to previous work \cite{stuerzlinger2006user,zhang2017interaction}. Alice notices that the location and size of UI elements on apps have only slightly changed, and that the overall appearance and structure of app UIs are still very similar (Figure \ref{fig:reflow-hands}).
Because of this, she is once again able to comfortably use her favorite apps, as she was previously able to do with her old phone, but she does not need to re-familiarize herself with the refined UIs.

\textbf{Customization:}
Alice may decide to customize the behavior of Reflow by opening its settings panel. In this panel, she may define lists of apps that she wishes the refinement feature to include or exclude. This could be useful for disabling the feature on apps that Alice already finds easy-to-use on her new smartphone. Similarly, Alice can define a shortcut (\textit{e.g.,} gesture or detected activity) to quickly toggle Reflow's functionality based on when it is useful. Finally, the settings panel allows her to reset or re-calibrate the feature to reflect updated preferences or physical affordances (for example, a smartphone case that makes it easier to grip the device).

\section{Related Work}
Several areas of related work have informed the design of Reflow: \textit{(i)} difficulties with touch interaction, \textit{(ii)} adaptive user interfaces, and \textit{(iii)} improving existing applications.

\subsection{Difficulties with Touch Interaction}
We first review literature related to the characterization of touch input on smartphones to identify why users like Alice experience interaction difficulty. Touch interaction can be analyzed using existing models of cursor-based selection~\cite{mackenzie1992fitts}, which has been extended to the touch screen~\cite{bi2013ffitts}. This type of analysis gives recommendations about the relative size and spacing of UI elements on touch-based apps.
Other work has focused on properties specific to touch interaction, such as the effect of the finger choice on selection accuracy. As examples, touch input is imprecise due to factors such as finger deformations (i.e., the ``fat finger" problem~\cite{vogel2007shift}), occlusions when the hand covers up UI elements~\cite{holz2010generalized,holz2011understanding}, and varying finger-to-screen ratios \cite{bergstrom2014modeling}. 

Many of these factors are spatially dependent (\textit{e.g.,} the finger occludes more of the screen when it makes contact with the touchscreen at an extreme angle).
Based on common hand postures used to grip a smartphone with one hand, \citet{le2019investigating} characterized the region of the phone that could be comfortably reached.
\citet{mayer2019finding} give further evidence through their analysis of 45 million touch events collected during touch-based gameplay, where they identified a region of the screen most likely to be comfortable to tap, known as the ``sweet spot.''
Based on this evidence, Reflow takes a spatially-dependent approach to personalizing touch interactions by modifying the position and size of UI elements.

\subsection{Adaptive User Interfaces}
Some UIs are constructed so that they can automatically reconfigure themselves dynamically depending on usage context. For example, adaptive UIs have been constructed to create alternate layouts for additional contexts~\cite{zeidler2017automatic} and provide accessibility benefits to older adults with motor and cognitive impairments~\cite{sarcar2018ability}.
This functionality can be manually programmed by app developers who expect usage difficulty to occur in certain contexts~\cite{kane2008getting, mariakakis2018drunk}.

Another approach is to define an objective function that measures how well a layout is perceived based on desired qualities~\cite{o2014learning, o2015designscape, swearngin2020scout}. SUPPLE is one example which uses this approach to automatically personalize an interface to facilitate faster access and lower error~\cite{gajos2010automatically}.
This approach has also shown encouraging results for optimizing application mockups for low-vision and motor impaired users \cite{gajos2008improving}.
In contrast to this approaches, Reflow uses an objective function that is learned from personalized usage data.
Specifically, we construct a neural network that predicts the time required to select on-screen elements based on performance on a calibration task, and our system aims to minimize this predicted value.

The approach by \citet{duan2020optimizing} employs gradient descent to optimize UIs with respect to estimated task completion time is most similar to our adaptation technique. In our work, we improve on this approach by: \textit{(i)} incorporating personalization through user-specific calibration data, \textit{(ii)} applying constrained optimization to minimize disruptions to the original UI, and \textit{(iii)} supporting end-to-end optimization of existing mobile app screens.
A limitation of many approaches is that they cannot be applied to real-world apps, as they often require the UI to be defined in a certain way (\textit{e.g.,} a layout definition) or implemented using specialized toolkits \cite{zeidler2013auckland, jiang2019orc,gajos2008improving, gajos2010automatically}. Reflow uses a pixel-based approach  without having to rely on the applications’ underlying implementation.

\subsection{Improving Existing Applications} %
Adaptive UIs that need to be built from scratch, using specific UI toolkits, and with specific requirements on developers, is severely limiting in practice. It is difficult to get developers to adopt any new UI framework, and this approach does not address the substantial legacy of existing applications not created to support this use.
An alternative is to repurpose and augment existing applications.
For example, existing UIs can be retargeted to support new modalities~\cite{swearngin2017genie} or support responsive resizing~\cite{jiang2021reverseorc, wu2021screen}.
\emph{Interaction proxies} apply runtime modification to existing mobile apps by ``repairing'' inaccessible or difficult-to-use UI elements \cite{zhang2017interaction}. Interaction proxies and other approaches that apply input/output redirection~\cite{stuerzlinger2006user} use specialized UI elements to re-render parts of the original UI while maintaining interactivity.

To support a broad range of existing applications by not having to rely on application dependencies or other application metadata, some approaches improve applications only from their pixels (visual appearance).
Template-based pixel matching has been used to locate icons or UI elements of interest on a screen to support early screen readers~\cite{outspoken} and end-user scripting~\cite{sikuli}.
Similarly, Prefab enables custom interaction techniques ({\em e.g.}, target-aware pointing) for desktop applications through pixel-based identification of user interface elements \cite{prefab}. An important contribution of the Reflow system is to demonstrate how UI refinements supporting better touch interaction can be made to existing mobile UIs directly from their pixels.

\section{Reflow}
\begin{figure*}[t!]
\centering
\includegraphics[width=\textwidth]{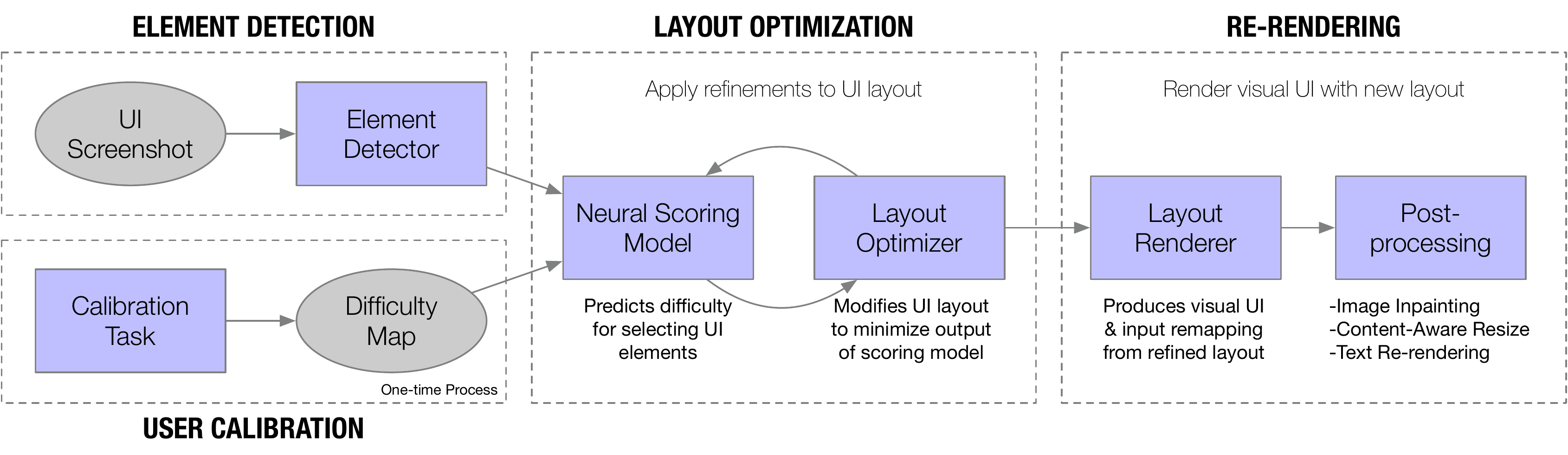}
\caption{A block diagram describing the architecture of Reflow. Reflow consists of 4 main stages: \textit{(i)} user calibration, \textit{(ii)} element detection, \textit{(iii)} layout optimization, and \textit{(iv)} re-rendering. Together, these stages allow an existing UI screenshot to be adapted to a personalized difficulty model.\label{fig:flow_diagram}}
\end{figure*}

Reflow is an end-to-end system that produces a refined UI from an original app's screenshot.
Reflow's operations occur in several stages (Figure \ref{fig:flow_diagram}).
First, the user is asked to perform a calibration task, which is used by Reflow to construct a difficulty map characterizing areas of the screen which may be hard-to-reach.
During runtime, an element detector is used to extract the layout of an existing app screen.
This layout is then optimized by repeatedly \textit{(i)} predicting the selection time for UI elements given the UI layout and user-specific difficulty map and \textit{(ii)} modifying the layout to minimize this predicted value while respecting stopping conditions designed to detect large disruptions (\textit{e.g.,} overlapping, other constraints).
Finally, the refined layout is re-rendered into a refined graphical user interface, where it can be presented to the user and made interactive.

\subsection{User Calibration}
\label{sec:calibration}

\subsubsection{Calibration Task}
To characterize the areas of the screen that are hard-to-reach for a user, we designed a one-handed calibration task that a user performs when first enabling the feature. We focused on one-handed touch interactions because these motor activities are more likely to cause selection challenges in user interfaces, particularly given the limited range and flexibility of the thumb~\cite{bergstrom2014modeling,chang2015understanding,negulescu2015grip}.
Users were asked to select on-screen targets, and the system recorded the \textit{(i)} selection time and \textit{(ii)} selection error (\textit{i.e.,} difference between the target position and recorded tap position).
These targets were uniformly spaced in a 4x8 grid, with a single, randomly selected target highlighted at a time.
After a target was selected, the next one was not immediately displayed; instead, it was delayed for the remainder of a timeout value.
This was done to allow the finger to return to a ``rest position'' after selection and reduce the influence of the previous target's location on the finger's starting position.
We set the timeout value to 3 seconds, based on our early observations and estimation of how long this process would take.

We conducted a data collection study with 10 participants (7M/2F/1 prefer not to disclose, ages 22-40, recruited within our organization) to characterize the input error for users and to initialize our system. We performed data collection remotely using video conference software.
Participants were asked to install an app on an iPhone, which was needed to run our software.
4/10 of our participants used an iPhone Xs, 2/10 used an iPhone 11, 3/10 used an iPhone Xs Max, and 1/10 used an iPhone 11 Pro.
Users were asked to hold their device with one hand and tap on targets using the same hand that they were holding the device with.

During the study, participants were asked to select targets placed at different locations on the screen.
In total, the study required less than 30 minutes and included both a practice and evaluation session.
9/10 participants held the phone in their right hand.

\subsubsection{Difficulty Map}
\begin{figure}[!]
\centering
\includegraphics[width=12pc]{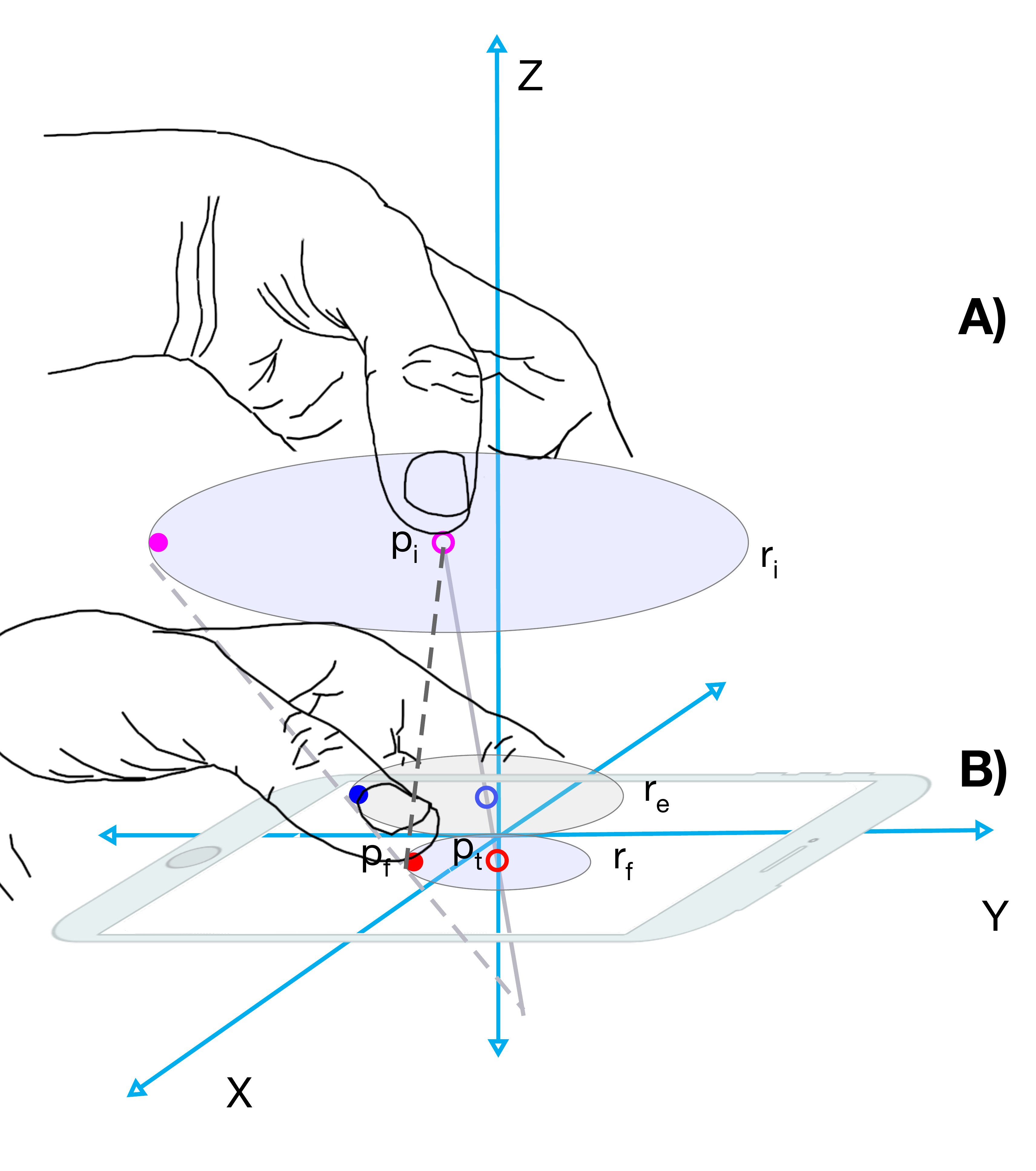}
\caption{A user taps a location on the screen by lowering a finger from a starting position $p_i$ (A) to a target location on the screen $p_t$ (B), whose path is shown as a dark gray dotted line. The circles represent region on the screen where the finger is expected to land given the current position. As the finger descends towards the screen, the circle shrinks, as the user ``hones in'' on the target location. The user's finger makes contact with the screen at $p_f$, which is a distance $r_f$ away from $p_t$ (B).}
\label{fig:projectivemodel}
\end{figure}
\begin{figure}[!]
\centering
\includegraphics[width=18pc]{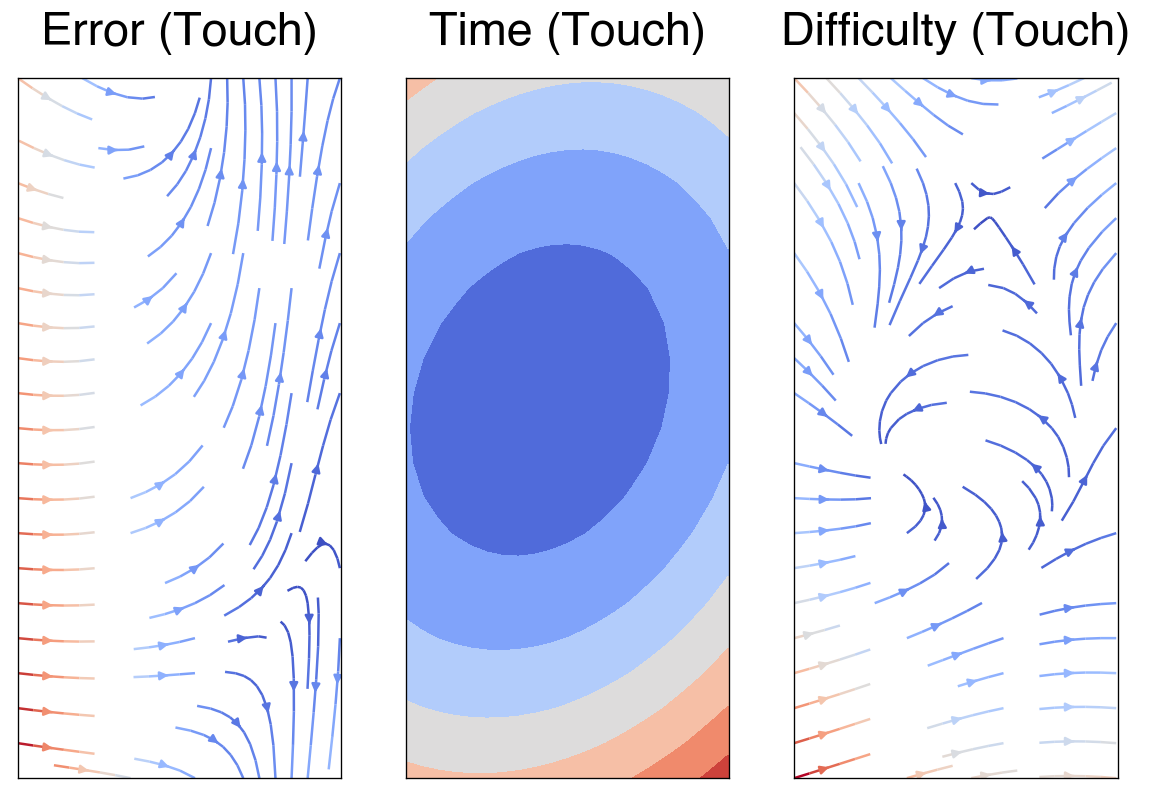}
\caption{An example of a difficulty map generated by our user calibration process. Data from the calibration task is first used to compute input error (Left) and selection time (Center) independently. We combine these two measurements to produce the ``adjusted error'', or difficulty map (Right).}
\label{fig:error_maps_cropped}
\end{figure}

Using calibration data from each user, we generated a personalized difficulty map~\cite{mott2019cluster} that estimates the relative difficulty of accessing any on-screen location  (including locations between two of the original calibration points).

We calculated difficulty by combining two measurements from the calibration data: \textit{(i)} input error and \textit{(ii)} selection time.
We introduce a procedure to ``normalize'' the raw error measurements by selection time.
Our procedure is based on the intuition that if a user selects two targets with equal accuracy (\textit{i.e.,} equal input error), the one that took longer to select (\textit{i.e.}, higher selection time) is more difficult.
We refer to this computed value as the ``adjusted error'', and it represents an estimation of input error given a constant selection time.
In other words, if a user was only given one second to select anywhere on the screen, what is the error we would expect for a given location?
Our approach is informed by prior work which suggests that selection-based interactions experience a tradeoff between speed and accuracy \cite{mackenzie1992fitts}.

Our computation of adjusted error is based on the standard Fitts's law equation.
Figure \ref{fig:projectivemodel} shows a user's finger, initially located at $p_i$ (\textit{i.e.,} resting position), selecting a target located at $p_t$, which is located a distance $A$ away.
\begin{equation}
    \begin{array}{l}
    t = a + b \cdot \log_2(A) \\
    t = a + b \cdot \log_2(||p_f - p_i||)
    \end{array}
    \label{eq:fittslawupdated}
\end{equation}
If the resting position $p_i$ is assumed to be constant for a single user (which previous research has shown is plausible \cite{le2018fingers,mayer2019finding}), we can estimate its location as a model parameter.
To do this---in addition to the standard Fitts' model parameters $a$ and $b$---we also learned $p_i$ by fitting Equation \ref{eq:fittslawupdated} to a dataset of pairs $\langle t_i, p_f \rangle$ using non-linear ordinary least squares.
The initial value of $p_i$ was set based on the resting location area identified in previous work \cite{le2018fingers}, and we constrained the position of $p_i$ so that it lies at most 5 inches above the device screen (a reasonable upper bound for the thumb's distance to the screen when holding a smartphone) and within the screen's x-y bounds.

To estimate the error magnitude $\mathbf{\epsilon}$ at a constant time $t_n$, we solve the following system of equations, where $a$ and $b$ are from the Fitts's model, and $\mathbf{d}$ is the distance the finger has traveled at $t_n$.
\begin{equation}
    (( A, r_f, 1 ) \times ( 0, r_i, 1 )) ( \mathbf{d}, \mathbf{\epsilon}, 1 )^{\mathbf{T}} = 0
\end{equation}
\begin{equation}
    \begin{array}{l}
    \mathbf{d} = 2^{\frac{t_n - a}{b}} \\
    \mathbf{\epsilon} = \frac{(r_f - r_i) \cdot \mathbf{d}}{A} + r_i
    \end{array}
\end{equation}
Using this procedure, we compute the adjusted error at each of the original calibration points (4x8 grid), then fit a bivariate polynomial (similar to a 2-D spline) to interpolate values in between.

To summarize, the output of the user calibration step is a function that returns the adjusted error for any location on the screen.
The adjusted error at location $(x, y)$ is an estimate of the offset of where a user's touch would land if given $t_n$ seconds to tap a target located at $(x, y)$.

\subsection{Element Detection}
In this stage, Reflow extracts the locations of on-screen UI elements using a CNN-based object detector and grouping heuristics~\cite{zhang2021screen}.
We performed additional post-processing on the output of the object detector to further improve performance.
First, we removed any detection that overlaps (by keeping the one with the higher confidence) or contains another (if a detection contains a text element, then the container is removed; otherwise, the children are removed).
The resulting layout contains no overlapping elements.
Because the downstream re-rendering stage (\cref{sec:rerendering}) involves cropping and moving image patches, the quality of element bounding boxes identified in element detection have affect the quality of the final output.
We employed a heuristic that refines the positions of bounding box edges by repeatedly \textit{(i)} computing the mean color of the pixels of each edge, \textit{(ii)} selecting the edge whose mean color is most dissimilar from the others, \textit{(iii)} adjusting its value by a small increment in the direction of improvement.

The output of the element detector is the UI layout of the original screenshot: $\Theta = \{\theta_1, \theta_2, \cdots ,\theta_n\}$ where $\theta_i = [x, y, w, h]$ describes the location and size of the $i$-th bounding box.
\subsection{Layout Optimization}

\subsubsection{Neural Scoring Model}
Using a difficulty map, we estimated the error a user would experience when selecting individual elements of the UI.
For a given app layout we define a scoring function $S(\Theta)$ which quantifies how likely the user is able to successfully select each element in the screen. 
We define the scoring function as
\begin{equation}
{\displaystyle S(\Theta) = \sum _{\theta \in \Theta}^{}\iint \limits _{R_{\theta}}P_\theta\,dA}
\label{eq:scoring_equation}
\end{equation}
where $P_\theta$ is the predicted distribution of where the user's finger will actually land when attempting to click the middle of an element $\theta$.
To estimate $P_\theta$, we centered 2-D Gaussian on the center of UI element $\theta$ and set its covariance based on the adjusted error at that location.
We compute $S(\Theta)$ using Monte Carlo integration.
For each UI element, $n$ samples are drawn from the 2-D Gaussian parameterized by the adjusted error at its center.
We set $n=30$ based on empirical observations of how many samples were needed for consistent results.
We scored a screen by counting the number of true positives over total number of points. 
\begin{figure}[t]
\centering
\hspace{-1.1pc}\includegraphics[width=0.95\columnwidth]{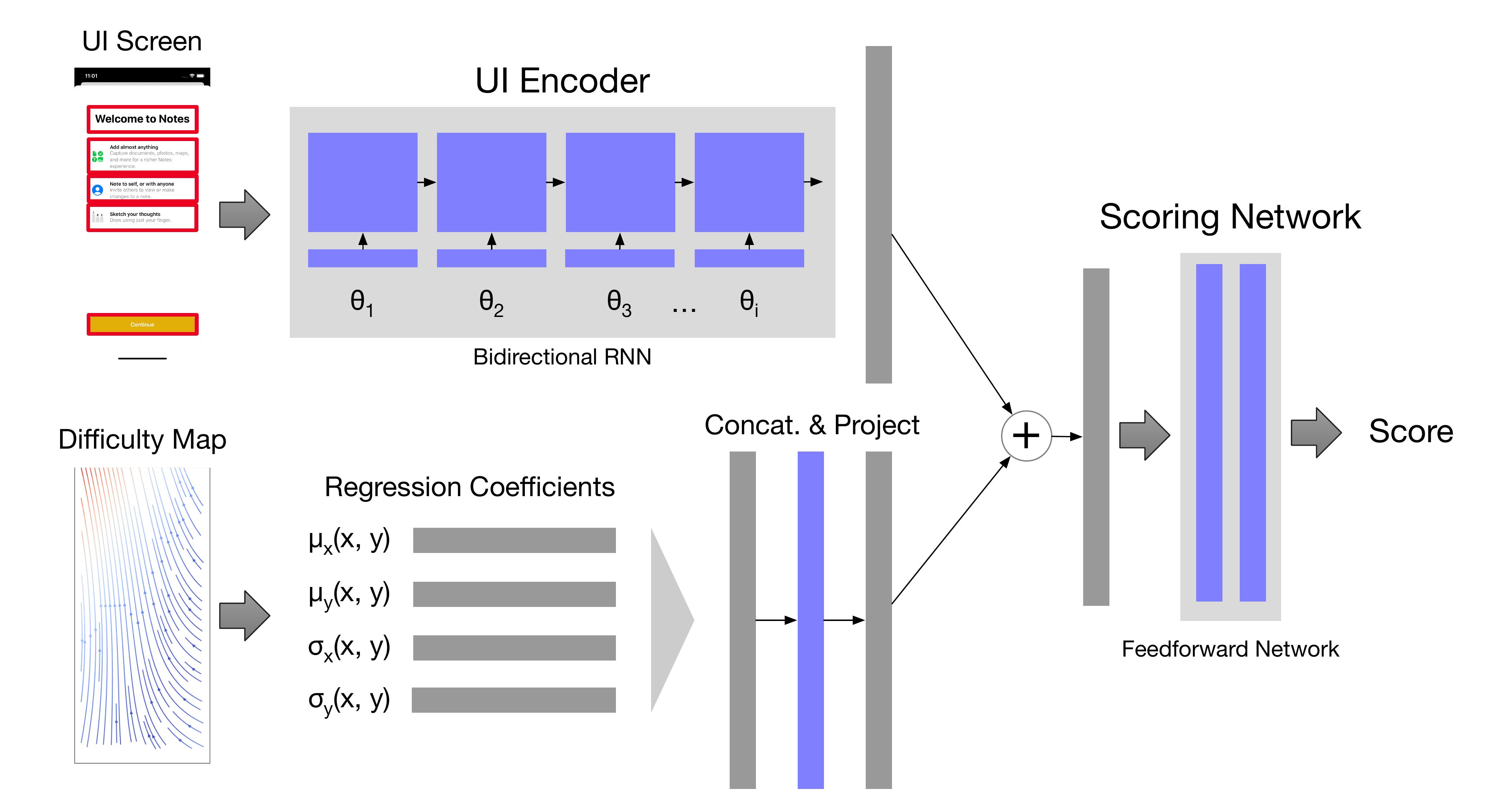}
\caption{The architecture for the neural scoring model used for scoring a UI screen given a spatial map of error. The network encodes the layout of UI elements using a bidirectional RNN and encodes the spatial difficulty map using the coefficients of a 2-D polynomial function fitted to the calibration points. These encoded representations are combined and fed into a feedforward network.}
\label{fig:scoringfunction}
\end{figure}

To further improve upon the speed and efficiency of our UI scoring function, we used a neural network to learn the result Monte Carlo scoring function.
Our model architecture (\cref{fig:scoringfunction}) consists of an LSTM-based UI layout encoder \cite{li2018predicting,duan2020optimizing} and a difficulty map encoder.
Because we parameterize the difficulty map as a model input, it allows for personalization at runtime---allowing re-calibration and removing the need to retrain the network for every new user.

We trained our neural scoring network on two datasets. 
The first was a large annotated dataset of 77,000 iOS app screens introduced in prior work \cite{zhang2021screen}.
The second dataset consisted of spatial difficulty maps collected from our user calibration dataset.
Because the second dataset is relatively small, we augmented it by introducing perturbations and randomly generated difficulty maps.
The network was trained by randomly selecting an application screen and a spatial difficulty map and computing the score using our Monte Carlo algorithm, which was used as the ground truth.
We trained our network using stochastic gradient descent until the validation loss stopped improving.

\subsubsection{Layout Optimization}
\begin{figure*}[t]
\centering
\includegraphics[width=0.85\textwidth]{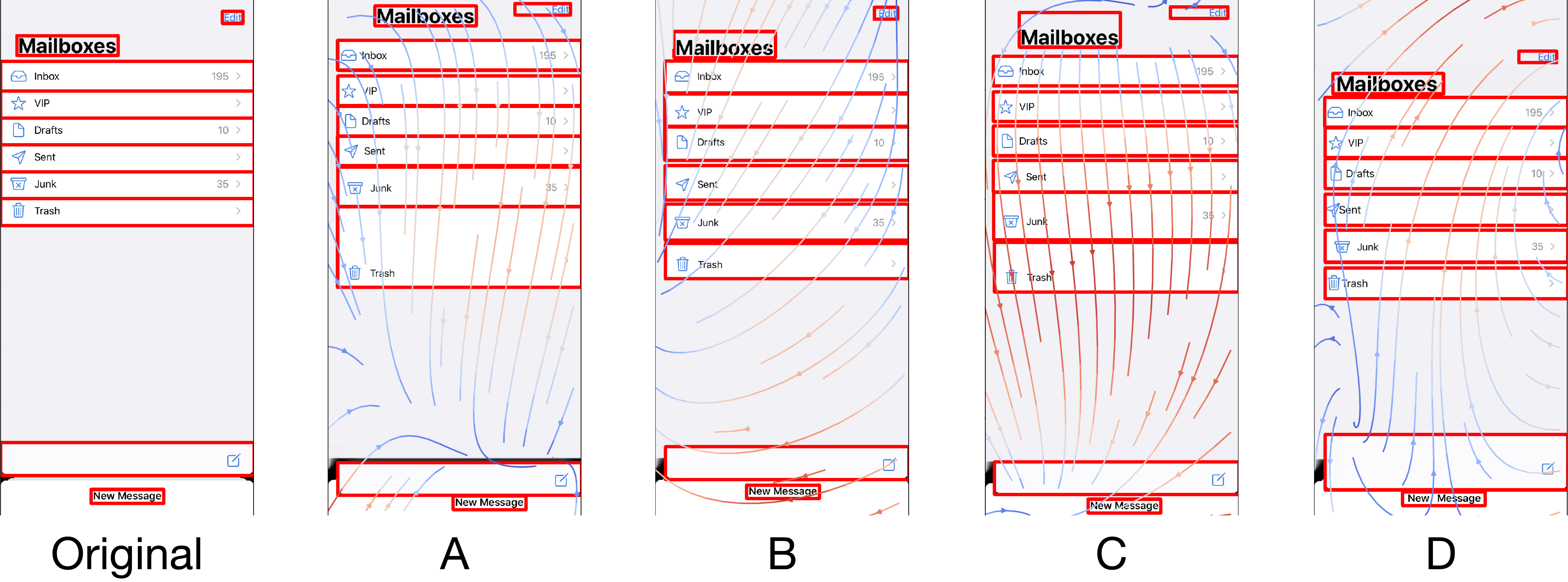}
\caption{An email app (``Original'') is optimized using four different spatial difficulty maps (``A'', ``B'', ``C'', ``D''). The result of the optimization process moves UI elements away from regions of higher relative difficulty (red) to lower relative difficulty (blue). Depending on the direction of error where an element is located, it may be stretched horizontally or vertically.}
\label{fig:example_gallery}
\end{figure*}
To optimize a UI layout, we use predictions from our neural scoring model to make modifications that reduce expected interaction difficulty.
One benefit of our model is that it is differentiable (since it is a neural network) and thus, given an initial screen layout and its corresponding score, we can use gradient-based optimizer \cite{kingma2014adam} to find an improved layout with a better score.
In service of our refinement approach to UI adaptation, we guide optimization and stop the process if disruptive changes are detected.

First, we add a regularization term that penalizes layouts that are proportionally dissimilar to the original.
This regularization term is defined as the cosine distance ($D_C$) between the pairwise $L_1$ displacements of each UI element ($\phi(\Theta)$).
This does not penalize the layout for increasing in size but attempts to maintains relationships between neighboring UI elements.

Next, we add ``corrective procedures'' after each optimization step to guide the optimization of certain properties (\textit{e.g.,} size and element).
We clamped element parameters to ensure they stay within a certain range and to prevent elements from becoming too small or large.
We also use a routine that detects overlapping regions and resolves them by moving overlapped elements further apart.
The overlap removal algorithm repeatedly tries to find intersecting region between pairs of UI elements, and if one exists, shifts them apart in the axis of least overlap.
This process is repeated until no more overlaps are detected or a max number of iterations is reached.
Despite our precaution, the final UI may still appear to contain overlaps due to inaccurate element detection (\textit{e.g.,} detection bounding box does not fully enclose element or includes multiple elements) and artifacts introduced by our re-rendering process.

Finally, we implemented an early stopping condition that is triggered when the overlap removal algorithm detects overlaps that are unresolvable, that is, overlaps that cannot be resolved by moving elements further apart.

Figure \ref{fig:example_gallery} shows examples of our algorithm's output for an email app optimized using different difficulty maps.

\subsection{Re-rendering}
\label{sec:rerendering}
Using the refined UI layout produced by layout optimization, we produce artifacts that are needed for end-user interaction: \textit{(i)} visual representation of the UI (UI screenshot) and \textit{(ii)} mapping between the original and refined UIs, needed for input redirection \cite{stuerzlinger2006user,zhang2017interaction}.
\subsubsection{Layout Renderer}
We use the refined UI layout and re-render the UI back into a visual representation.
First, Reflow outputs a mapping between regions of the original UI and the refined UI.
This mapping can be used to update the interactive regions of the UI using input/output redirection methods \cite{stuerzlinger2006user,zhang2017interaction} (\textit{e.g.,} clickable bounding box a button is updated to reflect its new optimized position).
To align the screen's visual appearance with the updated regions, image patches from the original screen are translated and resized to their new locations.

\subsubsection{Post-processing}
We use several post-processing strategies to preserve relevant aesthetic qualities (\textit{i.e.,} legibility of text).
We briefly describe three such techniques we used, which improve the legibility of: \textit{(i)} background areas, \textit{(ii)} text, and \textit{(iii)} image content.\newline

\begin{description}
\item[Image Inpainting.] 
Moving and resizing elements can result in ``holes'' so we employ an inpainting technique to generate visually plausible replacements.
We experimented with many different inpainting algorithms \cite{telea2004image,bertalmio2001navier,barnes2009patchmatch}, and found that most methods produced results of similar visual quality, since inpainted regions are unlikely to contain complex textures or structural features (most visual content is contained inside of the UI elements themselves).
Our implementation uses flow-based inpainting algorithm ~\cite{bertalmio2001navier} included in the OpenCV library \cite{opencv_library}.

\begin{figure}[!]
\centering
\includegraphics[width=0.95\linewidth]{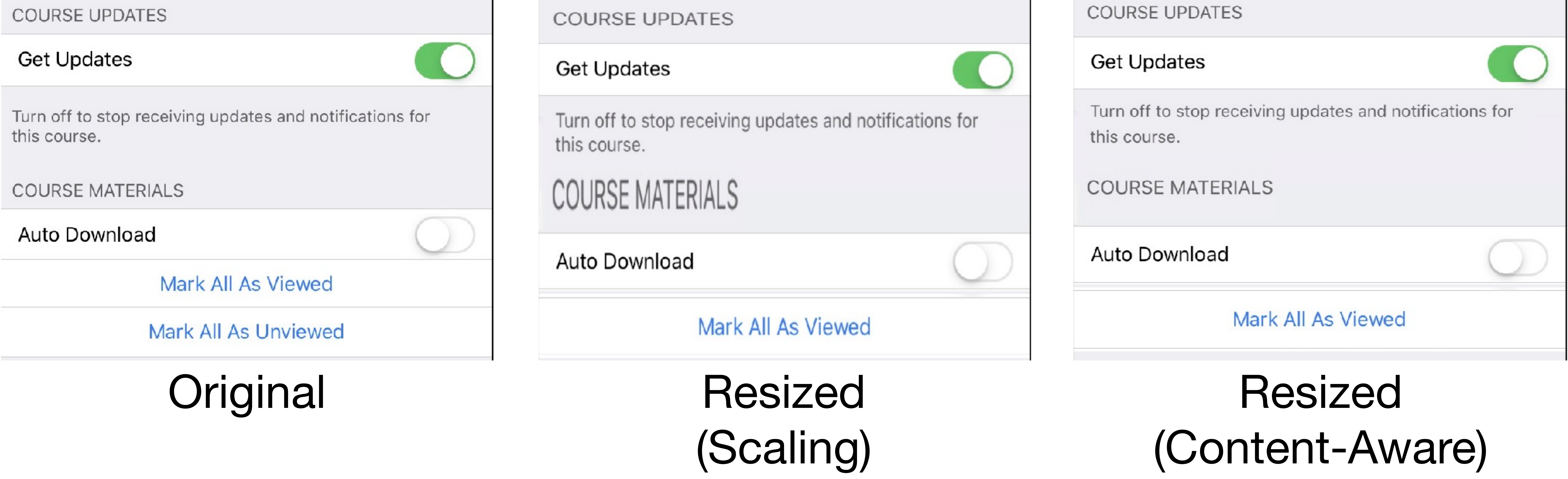}
\caption{An example of how content-aware resizing (Right) improves upon standard scaling (Center). Compared to the original screenshot (Left), scaling (Center) introduces distortions that make text less legible. Our approach (Right) preserves textual content.}
\label{fig:resize_example}
\end{figure}

\item[Content-Aware Resizing.] Because standard rescaling methods can distort elements (\cref{fig:resize_example}), our system uses an optimized method for resizing these regions.
We first create an image canvas with the target region's size.
The source image patch resized (retaining the original aspect ratio) to maximum size fits within the target dimensions.
Leftover space is filled using image inpainting.

\item[Text Re-rendering.] We explored a method specifically for resizing and replacing text (\textit{e.g.,} translation example application).
Our approach to resizing text involves detecting and extracting text using optical character recognition (OCR), re-rendering the text with the correct background and foreground colors, then inserting the result at the target location.
We used an off-the-shelf OCR system that recognizes text from images \cite{smith2007overview}. %
We estimated the original text's font size by rendering it using a known font and comparing its dimensions to the size of the original image patch.
Background and foreground color are extracted from the original image patch by performing k-means clustering on the pixels and extracting \textit{(i)} the largest cluster (background color), and \textit{(ii)} the cluster which is furthest away from the background color in pixel space (foreground color).
We created an image patch corresponding to the target dimensions and rendered the text with the correct foreground and background colors.

\end{description}

\subsection{Prototype Implementation}

For the purposes of our prototype, the Reflow system was implemented on a remote server. Screenshot images are sent from the iOS device to the remote server, which returns a re-rendered screenshot image that contained the UI refinements. For purposes of the user study (described next), the re-rendered screenshot is displayed to the user and touch events are recorded, which allowed us to conduct our user study. Prior work has demonstrated how such a re-rendered graphical UI could be used in a more advanced proxy setup to control the original underlying mobile app, which is how we imagine it would be used with existing applications in practice. We leave that implementation, and some difficult details ({\em e.g.}, handling scrolling GUIs) to future work.

\section{User Study}
\subsection{Procedure}

We empirically evaluated the performance of our system through a 45-minute user study.
We recruited 10 participants (5M/4F/1 Prefer not disclose, ages 24-36) within our organization.
Similar to how we collected data in the calibration task (\cref{sec:calibration}), we conducted our usability study remotely using a specialized data collection app and video conferencing software.
The breakdown of devices used by our participants were: 1/10 iPhone X, 5/10 iPhone Xs, 1/10 iPhone XR, 1/10 iPhone Xs Max, 1/10 iPhone 11, 1/10 iPhone 11 Pro.
8/10 participants held the phone using their right hand.

The usability study consisted of two phases, a calibration phase and a navigation phase, which included a practice session.
During calibration participants performed the user calibration task, which was used to generate personalized difficulty maps.
Participants were then given a short 5 minute break.
During the navigation phase, participants were presented with a series of app screens with a target UI element that was highlighted.
Participants were instructed to select the target element, which brought them to the next screen.

\begin{figure*}[t!]
\centering
\includegraphics[width=0.95\linewidth]{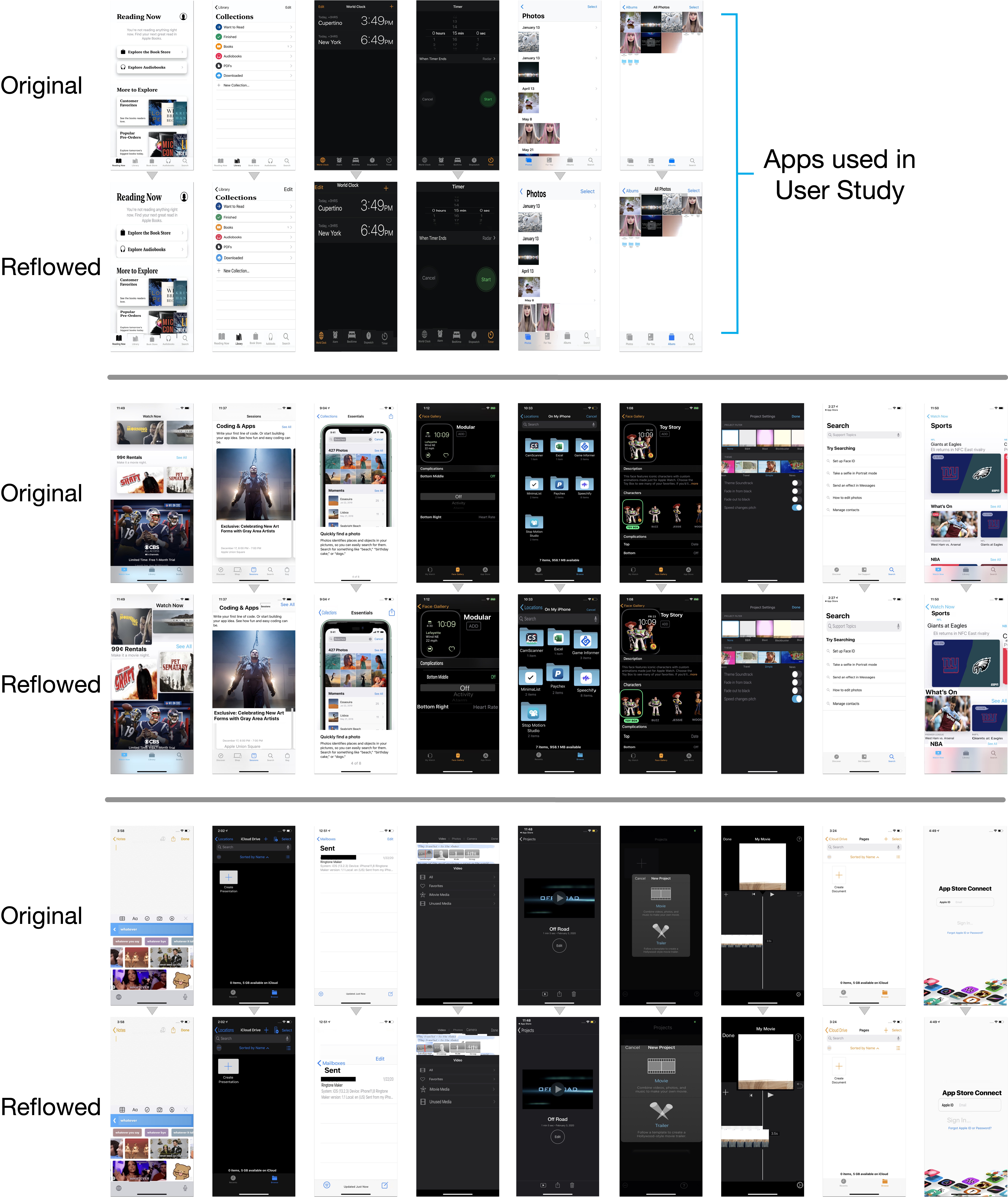}
\caption{A gallery of app screens optimized by Reflow. For the apps used in the User Study, we only show the first and last screens of each app (each app has 5 screens).
As was our intention, most changes by our system were conservative (resulting in little to no changes) and aimed at minimizing any negative effects introduced by drastic changes.
Some text (\textit{e.g.,} email address) is redacted in black.}
\label{fig:studyapps}
\end{figure*}

We chose 3 apps where users had to navigate through a total of 5 screens each (Figure \ref{fig:studyapps}).
3 apps were chosen to constrain the study design to fit under our time constraints, and we selected default applications on iOS that users are likely to be familiar with, since they come pre-installed.
All of these apps and screens were outside of the training set used to train our scoring model.
For each of the screens, we used our system (running on a remote server) to optimize its layout using the user's difficulty map generated during the calibration phase.

\subsection{Results}

\begin{table}[t!]
\begin{threeparttable}
\centering
\caption{Navigation Time Results from our User Study}
\begin{tabular}{lllll}
\toprule
        & \multicolumn{2}{l}{\textbf{Screen}} & \multicolumn{2}{l}{\textbf{App}}\\
        & Original     & Reflow      & Original    & Reflow     \\ \midrule
Books   & $1.1 \pm 0.4$    & $1.0 \pm 0.3$   & $4.2 \pm 1.0$   & $3.9 \pm 0.9$  \\
Clock   & $1.1 \pm 0.6$    & $1.0 \pm 0.5$   & $4.4 \pm 1.5$   & $3.8 \pm 1.3$  \\
Photos  & $1.2 \pm 0.5$    & $1.0 \pm 0.4$   & $4.7 \pm 1.4$   & $4.1 \pm 1.1$  \\ \midrule
Overall & $1.1 \pm 0.5$    & $1.0 \pm 0.4$   & $4.4 \pm 1.3$   & $4.0 \pm 1.1$\\ 
\bottomrule
\end{tabular}
\label{tab:usability_table}
\begin{tablenotes}
\item Navigation times from our user study. Values shown are the times ($M \pm SD$) needed to navigate a single screen. We report results for both per-screen and per-app navigation. Screen navigation refers to the time taken to advance one screen, while app navigation refers to the time taken to complete all screens for the app. For both measurements, Reflow's refinements resulted in 9\% faster navigation, on average.
\end{tablenotes}
\end{threeparttable}
\end{table}

Table \ref{tab:usability_table} shows the results of our user study.
We report both screen (time taken to navigate a single screen) and app (time taken to navigate all screens in an app) interaction times.
For both, Reflow's refinments resulted in an average of 9\% faster navigation.
This speed-up was significant for screen-level navigation ($T=2.76$, $p<0.01$, Cohen's $d=0.23$) and approached significance for app-level navigation ($T=1.99$, $p=0.06$, Cohen's $d=0.38$), due to small sample size for apps.
The small--medium effect size suggests that refinements lead to small, but significant improvements to interaction speed \cite{lakens2013calculating}.

\subsection{Post-study Improvements}
The results from our user study demonstrate the feasibility of our scoring model to optimize the layout of existing UIs.
Based on impromptu feedback from three participants during the study and our observations on how users interacted with Reflow, we made some improvements to further minimize the amount of disruption to the original UI.
Specifically, participants commented that the adaptation process affected the ordering of elements (\textit{e.g.,} an element was moved past an adjacent neighbor) and caused controls in UI structures navigation bars became mis-aligned.
We implemented additional heuristics to extract these constraints from the UI and used them further guide the optimization process.

\textbf{Constraint Extraction:}
Our approach maintains a list of constraints represented as equations and tests them against pairs of elements in the layout to determine if they hold.
We check for two types of constraints: \textit{(i)} relative positioning \textit{(ii)} alignment.

The enforcement of relative positioning ensures, for example, that if an element A was to the left of another element B, it remains to its left.
Relative positioning is represented using a multi constraint described by the following equation.
\begin{equation}
    x_i + w_i \leq x_j
\end{equation}
To reduce the number of relative constraints generated by our approach (making optimization more efficient), we remove redundant relationships through transitive reduction.
For example, if A is left of B and B is left of C, the relative positioning of A and C is implied and can be omitted as an explicit constraint.

Alignment constraints require that some part of two elements are arranged on the same line.
We consider 3 types of alignment between elements: \textit{(i)} beginning alignment, \textit{(ii)} center alignment, and \textit{(iii)} ending alignment.
These three types are computed for both the x and y dimension, resulting in a total of six possible element alignments.
Note that most interface builders may also allow mixing of these types \textit{e.g.,} the left edge of element A is aligned with the center of element B.
For simplicity, we omit these.

Because our UI element detector introduces a small amount of error in the bounding boxes, exact computation of alignment (\textit{e.g.,} computing $x_i = x_j$) would be inaccurate, leading to many undetected constraints.
Thus, we allow alignments to be met inexactly using slack variables.
We consider three types of alignment: beginning alignment, center alignment, and end alignment.
\begin{equation}
     \begin{array}{ll}
      x_i + \epsilon_i = x_j + \epsilon_j & \epsilon_i \geq 0, \epsilon_j \geq 0\\
      x_i + w_i + \epsilon_i = x_j + w_j + \epsilon_j & \epsilon_i \geq 0, \epsilon_j \geq 0\\
      x_i + \frac{w_i}{2} + \epsilon_i = x_j + \frac{w_j}{2} + \epsilon_j & \epsilon_i \geq 0, \epsilon_j \geq 0
    \end{array}   
\end{equation}

\textbf{Constrained Optimization:}
We used the non-convex optimization technique introduced by \citet{platt1988constrained} to augment our original layout optimizer.
Specifically, we added the heuristically extracted constraints as ``secondary functions'' of the main objective function (\textit{i.e.,} minimizing expected difficulty) that the optimizer aims to minimize.
Note that the while this technique allows the consideration of constraints, it does not guarantee that they will be met in the final solution.

\section{Heuristic Evaluation}
\label{sec:heuristic}

\subsection{Procedure}
To determine the acceptability of our improved system and to obtain qualitative feedback, we conducted a heuristic evaluation as described in \citet{10.5555/189200.189209}.
Heuristic evaluation has been previously been used as a design-centered analytic evaluation for user interfaces~\cite{mankoff2003heuristic,barik2016quick}. In a heuristic evaluation, expert evaluators examine the interface or aspects of the interaction to identify usability problems. \citet{10.5555/189200.189209} recommends the use of around three to five evaluators, so we contacted 10 people in our organization, in anticipation that some would be unavailable.
6 experts (3M, 3F) from different backgrounds (3 designers, 3 accessibility experts) agreed to participate and provided diverse feedback.
Recruitment was done via convenience sampling---we reached out to potential evaluators using our organization's messaging software.
All of our evaluators had multiple years of experience in their field and five had doctoral degrees in their respective areas of specialization.

The heuristic evaluation was conducted online, and evaluators were sent a link to the evaluation materials.
The link first gave a brief description of Reflow and the evaluation task.
Evaluators were instructed to watch a video clip that showed four usage scenarios (\textit{e.g.,} while walking or holding a shopping bag with one hand) where a user interacted with app UIs.
Original and adapted versions of the UI were overlayed on the video and allowed evaluators to assess different qualities of the UI in context.
Evaluators could pause and replay the video as many times as they needed during the evaluation to more closely inspect aspects of the UI and usage scenario.

We provided a questionnaire with a set of UI heuristics for them to evaluate, informed by prior work on layout usability \cite{shiripour2021grid, todi2016sketchplore}.
We removed heuristics (\textit{e.g.,} color harmony) that were not applicable to Reflow and settled on the following: alignment, selection time, visual clutter, saliency, and element grouping.
For each heuristic, we provided a brief description and asked evaluators to rate adherence using a 5-point Likert scale for comparison (``Significantly Worse (1)'' to ``Significantly Better (5)'').
We also asked evaluators to provide rationale for their ratings and encouraged elaboration on the positive and negative aspects on the layout changes introduced by Reflow.

\subsection{Results}

\input{table_heuristic}

\cref{tab:heuristic_eval} shows each evaluator's 5-point Likert-type item responses. Overall, the results of the heuristic evaluation confirm that Reflow improves touch efficiency while minimally disrupting the user experience.
We summarize the feedback from our evaluators along each of the heuristics.

\textbf{Alignment:}
\textit{Heuristic.}
Alignment refers to the internal alignment of elements with each other.
\textit{Evaluation.}
Most evaluators agreed that Reflow's refinements slightly impacted alignment.
While our improved layout optimizer does detect and account for UI constraints such as alignment, it does not guarantee that these constraints are met.
We found that some evaluators' intepretations of ``alignment'' included aspects such as spacing (not explicitly handled by our system) as well as edge alignment (handled by our system) ($\mathrm{E3_{AX}}$).
Nevertheless, $\mathrm{E2_{AX}}$ observed that some layouts \textit{``probably could change more without hurting understanding,''} indicating that Reflow's refinements kept semantic relations are kept intact.

\textbf{Selection Time:}
\textit{Heuristic.}
Selection time refers to the time needed to select the target element.
Since the evaluators did not directly operate each UI, we asked evaluators to estimate the selection time from the relative positioning of elements (\textit{i.e.,}), as in previous work \cite{todi2016sketchplore,shiripour2021grid}.
\textit{Evaluation.}
Feedback from our evaluators mostly indicated that our adaptations allow for more rapid selection times during use, but evaluators expressed some uncertainty as they were estimating purely from visual appearance ($\mathrm{E1_{AX}}$, $\mathrm{E3_{AX}}$).
We did indeed empirically measure selection times from our user study (\cref{tab:usability_table}), and our expert's estimates are consistent with our findings.
$\mathrm{E5_{DS}}$ pointed out that our system allowed faster interaction \textit{``without having to use reachability''} and would \textit{``certainly save time,''} since performing the Reachability gesture (\cref{sec:example_scenario}) would require more time than using Reflow.

Taken together with other feedback (comment by $\mathrm{E2_{AX}}$ on alignment), this suggests that supporting manual adjustment of optimization levels may help Reflow better serve users with differing goals---either prioritizing selection time or preserving aspects of the original UI (\textit{e.g.,} alignment).

\textbf{Visual Clutter:}
\textit{Heuristic.}
Visual clutter refers to how confusing a display is.
The more cluttered a display, the more difficult it is for an element to catch a user's attention
\textit{Evaluation.}
$\mathrm{E2_{AX}}$ pointed out that movement of semantically-important elements contributed to confusion.
There was some level of disagreement between evaluators ($\mathrm{E3_{AX}}$, $\mathrm{E5_{DS}}$), which was in part influenced by different interpretations of the heuristic.
For example, $\mathrm{E3_{AX}}$ attributed the higher levels of visual clutter to \textit{``worse alignment, unexpected/unintuitive negative space, and overlapping elements,''} indicating the consideration of multiple factors such as alignment, spacing, and overlap.

\textbf{Saliency:}
\textit{Heuristic.}
Saliency refers to the degree to which important elements are more likely to catch the attention of the user.
\textit{Evaluation.}
Evaluators suggested that our system's modifications did not induce significant changes in saliency, which is desirable since a large saliency change would imply a violation of design intent.
$\mathrm{E3_{AX}}$ indicated that the refinement process only resulted in \textit{``subtle changes''} and did not make a single element any more or less salient than the others.
$\mathrm{E2_{AX}}$ noted that size refinements, which allows elements in ``difficult areas'' to be selected more easily, positively impacted the saliency of small items by enlarging their tap targets; however the same behavior was viewed negatively by $\mathrm{E1_{AX}}$: \textit{``the adaptations didn't have much impact on saliency, so I almost selected about the same, but I noticed that the tabs have increased size so they may be over-emphasized.''}

\textbf{Element Grouping:}
\textit{Heuristic.}
Elements that are clustered near each other are perceived as a unified object.
\textit{Evaluation.}
Element grouping was also minimally affected by Reflow's refinement process.
While evaluators generally agreed that element grouping remained ``about the same'' ($\mathrm{E2_{AX}}$, $\mathrm{E4_{DS}}$, $\mathrm{E6_{DS}}$), some pointed out instances where the adjustment of element spacing led to ambiguity: \textit{``on the 4th screenshot pair, the inbox and arrows have scrunched down too closely to the App Store and To field, leaving not enough separation''} ($\mathrm{E1_{AX}}$).

\section{Discussion}

\subsection{Flexible Personalization through Difficulty Maps}

An architectural decision choice we made in Reflow was to parameterize the difficultly map as an explicit input to the neural scoring model (\cref{fig:flow_diagram}). Doing so has a number of advantages.

Difficulty maps are the key mechanism through which we enable personalization, because they allow users to recalibrate their touch interactions without having to retrain the neural scoring model for every user. Having to retrain a model for every user is not practical, especially if the retraining must be done on a mobile device.

An second advantage of parameterizing difficulty maps is that Reflow can be easily extended to other types of interactions. While our studies evaluated one-handed touch interactions, the Reflow is not restricted to this. Difficulty maps can also be constructed to support reachability, handedness, motor impairments, or other touch accommodations needed to support users. Difficulty maps can even go beyond touch interactions, for example, when using a pen input or other pointing device. Once a mode of interaction is translated to a difficulty map, the rest of the Reflow architecture can take advantage of this---without requiring any changes the remaining stages in the Reflow pipeline.

Finally, difficulty maps make it possible to offer users a set of pre-defined profiles for common touch interaction scenarios. Pre-defined maps could cover scenarios such as one-handed usage and allow many users to benefit from adaptation without having to perform manual calibration. Should these pre-defined maps not support the user, they can always fallback to performing a quick calibration to construct a personalized difficulty map.

\subsection{Design Space between Touch Efficiency and Layout Preservation}

Reflow applies \emph{refinements}---small UI adaptations that minimally disrupt the UI layout---as a design choice for performing layout optimization. Our heuristic evaluation validates that this decision choice as an appropriate one. However, this decision choice is only one possible point across the full design space, which includes a spectrum of trade-offs between improving touch efficiency against preserving the existing layout. Users may have different preferences along this design space.

For example, all evaluators in our heuristic evaluation---except E4---indicated that selection time would be improved by Reflow. E4 desired to see more dramatic improvements to touch interactions, even if this would require Reflow to make more substantial disruptions to the layout. In contrast, E3 disliked the clustering that resulted from Reflow’s layout optimization. Users such as E3 may find it acceptable to have fewer touch improvements, if this would result in smaller layout disruptions.

One possibility is to give users more control over this design space. Consider a slider control that allows the user to select points in the design space between maximizing touch interaction efficiency, minimizing layout disruption, or some balance in between these two extremes. Users may also want to selectively enable or disable specific types of optimizations, for example, they may prefer not to allow the size of UI elements to change. Users may even have different preferences between applications, with varying expectations about layout disruption for the different applications.

\subsection{Opportunities for Reflow}
Through our prototype implementation and experiments, we identified opportunities and future work for Reflow, including \textit{(i)} extracting and incorporating screen semantics, \textit{(ii)} improving interaction fidelity, and \textit{(iii)} studying the effect of applying refinements over time.

\subsubsection{Extracting Screen Semantics}
Several of these opportunities involve addressing existing limitations of Reflow. First, the set of constraints that we automatically extract should be extended to support those available in conventional UI authoring tools, such as vertical and horizontal alignment guides, distributing vertical and horizontal spacing, and resizing elements across axes. Including these constraints and inference techniques~\cite{lutteroth2008automated,bielik2018robust,jiang2021reverseorc, krosnick2018expresso} in Reflow would further minimize layout disruptions. Currently, Reflow internally represents UIs as a list of bounding boxes, but the limitations of this approach are that it is unable to capture semantic relationships \emph{between} elements during inference. Because of this restriction, Reflow currently treats a list of $n$ UI elements as $n$ unrelated UI elements, which is potentially a lost opportunity for layout optimization.

\subsubsection{Improved Interaction Fidelity}
Reflow's pixel-based pipeline consumes an image (\textit{i.e.,} screenshot) of the current UI as input and also generates an image as output.
The refined UI is made interactive by making certain parts of the output image respond to touch events, which can then be forwarded to the original app.
While this allows Reflow to be applied to any screen, it may lead to poor performance on screens with dynamic properties such as animated content and scrolling because this behavior cannot be adequately captured in a static screenshot.

One way to extend the current approach to handle these cases is to perform this image-to-image process multiple times per second, thus updating the output as frequently as dynamic behavior occurs.
Besides the optimization and performance challenges this entails, repeatedly optimizing single video frames as independent inputs may lead to artifacts such as jitter.
An alternative approach is to regenerate the interface from extracted semantics and interfaces, which has previously been applied to web applications \cite{nichols2008highlight}.
Recent work in pixel-based semantic extraction \cite{fischer2018brassau,wu2021screen} suggests this may also be possible for mobile UIs.

\subsubsection{Extended Evaluation}
Finally, a longer-term usage study may reveal more detailed effects of adaptive UIs and, more specifically, our refinements approach.
Although our studies demonstrate the effectiveness of refinements for improving touch efficiency, it would be important to evaluate Reflow in longitudinal studies. The benefits of Reflow can only be fully realized through cumulative use: the longer users use Reflow, the more opportunities they have to take advantage of the adapted UIs. As \citet{gray2000milliseconds} observe---``milliseconds matter’’---and even seemingly small improvements add up over frequent and repeated touch interactions.

\section{Conclusion}
In this paper, we introduced Reflow, a system that automatically applies small, personalized UI adaptations---called \emph{refinements}---to mobile app screens to improve touch efficiency.
Reflow supports real-world UIs without any source code or metadata dependencies through pixed-based element detection. Reflow optimizes a UI by \textit{(i)} extracting its layout from its screenshot, \textit{(ii)} refining its layout, and \textit{(iii)} re-rendering the UI to reflect these modifications. 
We conducted a user study with 10 participants and found that UIs optimized by Reflow were on average 9\% faster to use.
We conducted a heuristic evaluation with 6 experts to elicit qualitative feedback about Reflow and validate that the system's UI refinements improve selection time while minimizing layout disruption.
The results of our work demonstrate that refinements applied by Reflow are a useful UI adaptation technique to improve touch interactions.

\bibliographystyle{ACM-Reference-Format}
\bibliography{sample-base}

\end{document}

%% file: table_heuristic.tex
\newcommand{\heplusone}{\cellcolor{blue!10}$+1$}
\newcommand{\heplustwo}{\cellcolor{blue!20}$+2$}
\newcommand{\heplusthree}{\cellcolor{blue!30}$+3$}
\newcommand{\heplusfour}{\cellcolor{blue!40}$+4$}

\newcommand{\hezero}{\cellcolor{gray!5}}
\newcommand{\heblank}{\cellcolor{white}$\cdot$}

\newcommand{\heminusone}{\cellcolor{orange!10}$-1$}
\newcommand{\heminustwo}{\cellcolor{orange!20}$-2$}
\newcommand{\heminusthree}{\cellcolor{orange!30}$-3$}
\newcommand{\heminusfour}{\cellcolor{orange!40}$-4$}

\begin{table}[t]
\centering
\begin{threeparttable}
\caption{Relative Likert Scores for Heuristic Evaluation}
\begin{tabulary}{\linewidth}{Lcccccc}
\toprule
& \multicolumn{3}{c}{AX} & \multicolumn{3}{c}{DS} \\
\cmidrule(lr){2-4} \cmidrule(lr){5-7}
Heuristic & E1 & E2 & E3 & E4 & E5 & E6\\
\midrule
\textsc{Alignment} & 
  \heminusone & 
  \hezero & 
  \heminusone & 
  \heminusone & %
  \heplusone & 
  \heminusone
\\
\textsc{Selection Time} &
  \heplusone &
  \heplusone &
  \heplusone &
  \hezero &
  \heplusone &
  \heplusone
\\
\textsc{Clutter} &
  \hezero &
  \heminusone & 
  \heminustwo &
  \heminusone & 
  \heplustwo &
  \heminusone 
\\
\textsc{Saliency} &
  \heminusone &
  \heplusone & 
  \hezero &
  \heminusone & 
  \heplusone &
  \heminusone 
\\
\textsc{Grouping} &
  \heminusone &
  \hezero & 
  \heminusone &
  \hezero & 
  \heplusone &
  \hezero
\\

\bottomrule
\end{tabulary}
\label{tab:heuristic_eval}
\begin{tablenotes}
\item Expert evaluations from heuristic evaluation. Ratings are normalized to show deviation from the neutral option (``About the same (3)"). Positive scores indicate better change and negative scores indicate worse change. AX denotes accessibility expert while DS denotes design expert.
\end{tablenotes}
\end{threeparttable}
\end{table}

%% file: sample-acmsmall.bbl

\begin{thebibliography}{57}

%%% the Web address.

\ifx \showCODEN    \undefined \def \showCODEN     #1{\unskip}     \fi
\ifx \showDOI      \undefined \def \showDOI       #1{#1}\fi
\ifx \showISBNx    \undefined \def \showISBNx     #1{\unskip}     \fi
\ifx \showISBNxiii \undefined \def \showISBNxiii  #1{\unskip}     \fi
\ifx \showISSN     \undefined \def \showISSN      #1{\unskip}     \fi
\ifx \showLCCN     \undefined \def \showLCCN      #1{\unskip}     \fi
\ifx \shownote     \undefined \def \shownote      #1{#1}          \fi
\ifx \showarticletitle \undefined \def \showarticletitle #1{#1}   \fi
\ifx \showURL      \undefined \def \showURL       {\relax}        \fi
% invisible to TeX
\providecommand\bibfield[2]{#2}
\providecommand\bibinfo[2]{#2}
\providecommand\natexlab[1]{#1}
\providecommand\showeprint[2][]{arXiv:#2}

\bibitem[Barik et~al\mbox{.}(2016)]%
        {barik2016quick}
\bibfield{author}{\bibinfo{person}{Titus Barik}, \bibinfo{person}{Yoonki Song},
  \bibinfo{person}{Brittany Johnson}, {and} \bibinfo{person}{Emerson
  Murphy-Hill}.} \bibinfo{year}{2016}\natexlab{}.
\newblock \showarticletitle{From quick fixes to slow fixes: Reimagining static
  analysis resolutions to enable design space exploration}. In
  \bibinfo{booktitle}{\emph{2016 IEEE International Conference on Software
  Maintenance and Evolution (ICSME)}}. IEEE, \bibinfo{pages}{211--221}.
\newblock


\bibitem[Barnes et~al\mbox{.}(2009)]%
        {barnes2009patchmatch}
\bibfield{author}{\bibinfo{person}{Connelly Barnes}, \bibinfo{person}{Eli
  Shechtman}, \bibinfo{person}{Adam Finkelstein}, {and} \bibinfo{person}{Dan~B
  Goldman}.} \bibinfo{year}{2009}\natexlab{}.
\newblock \showarticletitle{PatchMatch: A randomized correspondence algorithm
  for structural image editing}. In \bibinfo{booktitle}{\emph{ACM Transactions
  on Graphics (ToG)}}, Vol.~\bibinfo{volume}{28}. ACM, \bibinfo{pages}{24}.
\newblock


\bibitem[Bergstrom-Lehtovirta and Oulasvirta(2014)]%
        {bergstrom2014modeling}
\bibfield{author}{\bibinfo{person}{Joanna Bergstrom-Lehtovirta} {and}
  \bibinfo{person}{Antti Oulasvirta}.} \bibinfo{year}{2014}\natexlab{}.
\newblock \showarticletitle{Modeling the functional area of the thumb on mobile
  touchscreen surfaces}. In \bibinfo{booktitle}{\emph{Proceedings of the SIGCHI
  Conference on Human Factors in Computing Systems}}.
  \bibinfo{pages}{1991--2000}.
\newblock


\bibitem[Bertalmio et~al\mbox{.}(2001)]%
        {bertalmio2001navier}
\bibfield{author}{\bibinfo{person}{Marcelo Bertalmio},
  \bibinfo{person}{Andrea~L Bertozzi}, {and} \bibinfo{person}{Guillermo
  Sapiro}.} \bibinfo{year}{2001}\natexlab{}.
\newblock \showarticletitle{Navier-stokes, fluid dynamics, and image and video
  inpainting}. In \bibinfo{booktitle}{\emph{Proceedings of the 2001 IEEE
  Computer Society Conference on Computer Vision and Pattern Recognition. CVPR
  2001}}, Vol.~\bibinfo{volume}{1}. IEEE, \bibinfo{pages}{I--I}.
\newblock


\bibitem[Bi et~al\mbox{.}(2013)]%
        {bi2013ffitts}
\bibfield{author}{\bibinfo{person}{Xiaojun Bi}, \bibinfo{person}{Yang Li},
  {and} \bibinfo{person}{Shumin Zhai}.} \bibinfo{year}{2013}\natexlab{}.
\newblock \showarticletitle{FFitts law: modeling finger touch with fitts' law}.
  In \bibinfo{booktitle}{\emph{Proceedings of the SIGCHI Conference on Human
  Factors in Computing Systems}}. \bibinfo{pages}{1363--1372}.
\newblock


\bibitem[Bielik et~al\mbox{.}(2018)]%
        {bielik2018robust}
\bibfield{author}{\bibinfo{person}{Pavol Bielik}, \bibinfo{person}{Marc
  Fischer}, {and} \bibinfo{person}{Martin Vechev}.}
  \bibinfo{year}{2018}\natexlab{}.
\newblock \showarticletitle{Robust relational layout synthesis from examples
  for Android}.
\newblock \bibinfo{journal}{\emph{Proceedings of the ACM on Programming
  Languages}} \bibinfo{volume}{2}, \bibinfo{number}{OOPSLA}
  (\bibinfo{year}{2018}), \bibinfo{pages}{1--29}.
\newblock


\bibitem[Bigham(2014)]%
        {opportunistic}
\bibfield{author}{\bibinfo{person}{Jeffrey~P. Bigham}.}
  \bibinfo{year}{2014}\natexlab{}.
\newblock \showarticletitle{Making the Web Easier to See with Opportunistic
  Accessibility Improvement}. In \bibinfo{booktitle}{\emph{Proceedings of the
  27th Annual ACM Symposium on User Interface Software and Technology}}
  (Honolulu, Hawaii, USA) \emph{(\bibinfo{series}{UIST ’14})}.
  \bibinfo{publisher}{Association for Computing Machinery},
  \bibinfo{address}{New York, NY, USA}, \bibinfo{pages}{117–122}.
\newblock
\showISBNx{9781450330695}
\urldef\tempurl%
\url{https://doi.org/10.1145/2642918.2647357}
\showDOI{\tempurl}


\bibitem[Bradski(2000)]%
        {opencv_library}
\bibfield{author}{\bibinfo{person}{G. Bradski}.}
  \bibinfo{year}{2000}\natexlab{}.
\newblock \showarticletitle{{The OpenCV Library}}.
\newblock \bibinfo{journal}{\emph{Dr. Dobb's Journal of Software Tools}}
  (\bibinfo{year}{2000}).
\newblock


\bibitem[Chang et~al\mbox{.}(2015)]%
        {chang2015understanding}
\bibfield{author}{\bibinfo{person}{Youli Chang}, \bibinfo{person}{Sehi L'Yi},
  \bibinfo{person}{Kyle Koh}, {and} \bibinfo{person}{Jinwook Seo}.}
  \bibinfo{year}{2015}\natexlab{}.
\newblock \showarticletitle{Understanding users' touch behavior on large mobile
  touch-screens and assisted targeting by tilting gesture}. In
  \bibinfo{booktitle}{\emph{Proceedings of the 33rd Annual ACM Conference on
  Human Factors in Computing Systems}}. \bibinfo{pages}{1499--1508}.
\newblock


\bibitem[Dixon and Fogarty(2010)]%
        {prefab}
\bibfield{author}{\bibinfo{person}{Morgan Dixon} {and} \bibinfo{person}{James
  Fogarty}.} \bibinfo{year}{2010}\natexlab{}.
\newblock \showarticletitle{Prefab: Implementing Advanced Behaviors Using
  Pixel-Based Reverse Engineering of Interface Structure}. In
  \bibinfo{booktitle}{\emph{Proceedings of the SIGCHI Conference on Human
  Factors in Computing Systems}} (Atlanta, Georgia, USA)
  \emph{(\bibinfo{series}{CHI ’10})}. \bibinfo{publisher}{Association for
  Computing Machinery}, \bibinfo{address}{New York, NY, USA},
  \bibinfo{pages}{1525–1534}.
\newblock
\showISBNx{9781605589299}
\urldef\tempurl%
\url{https://doi.org/10.1145/1753326.1753554}
\showDOI{\tempurl}


\bibitem[Duan et~al\mbox{.}(2020)]%
        {duan2020optimizing}
\bibfield{author}{\bibinfo{person}{Peitong Duan}, \bibinfo{person}{Casimir
  Wierzynski}, {and} \bibinfo{person}{Lama Nachman}.}
  \bibinfo{year}{2020}\natexlab{}.
\newblock \showarticletitle{Optimizing user interface layouts via gradient
  descent}. In \bibinfo{booktitle}{\emph{Proceedings of the 2020 CHI Conference
  on Human Factors in Computing Systems}}. \bibinfo{pages}{1--12}.
\newblock


\bibitem[Eng et~al\mbox{.}(2006)]%
        {eng2006generating}
\bibfield{author}{\bibinfo{person}{Katherine Eng}, \bibinfo{person}{Richard~L
  Lewis}, \bibinfo{person}{Irene Tollinger}, \bibinfo{person}{Alina Chu},
  \bibinfo{person}{Andrew Howes}, {and} \bibinfo{person}{Alonso Vera}.}
  \bibinfo{year}{2006}\natexlab{}.
\newblock \showarticletitle{Generating automated predictions of behavior
  strategically adapted to specific performance objectives}. In
  \bibinfo{booktitle}{\emph{Proceedings of the sigchi conference on human
  factors in computing systems}}. \bibinfo{pages}{621--630}.
\newblock


\bibitem[Everett and Byrne(2004)]%
        {everett2004unintended}
\bibfield{author}{\bibinfo{person}{Sarah~P Everett} {and}
  \bibinfo{person}{Michael~D Byrne}.} \bibinfo{year}{2004}\natexlab{}.
\newblock \showarticletitle{Unintended effects: Varying icon spacing changes
  users' visual search strategy}. In \bibinfo{booktitle}{\emph{Proceedings of
  the SIGCHI conference on Human factors in computing systems}}.
  \bibinfo{pages}{695--702}.
\newblock


\bibitem[Fischer et~al\mbox{.}(2018)]%
        {fischer2018brassau}
\bibfield{author}{\bibinfo{person}{Michael Fischer}, \bibinfo{person}{Giovanni
  Campagna}, \bibinfo{person}{Silei Xu}, {and} \bibinfo{person}{Monica~S Lam}.}
  \bibinfo{year}{2018}\natexlab{}.
\newblock \showarticletitle{Brassau: automatic generation of graphical user
  interfaces for virtual assistants}. In \bibinfo{booktitle}{\emph{Proceedings
  of the 20th International Conference on Human-Computer Interaction with
  Mobile Devices and Services}}. \bibinfo{pages}{1--12}.
\newblock


\bibitem[Gajos et~al\mbox{.}(2010)]%
        {gajos2010automatically}
\bibfield{author}{\bibinfo{person}{Krzysztof~Z Gajos},
  \bibinfo{person}{Daniel~S Weld}, {and} \bibinfo{person}{Jacob~O Wobbrock}.}
  \bibinfo{year}{2010}\natexlab{}.
\newblock \showarticletitle{Automatically generating personalized user
  interfaces with Supple}.
\newblock \bibinfo{journal}{\emph{Artificial Intelligence}}
  \bibinfo{volume}{174}, \bibinfo{number}{12-13} (\bibinfo{year}{2010}),
  \bibinfo{pages}{910--950}.
\newblock


\bibitem[Gajos et~al\mbox{.}(2008)]%
        {gajos2008improving}
\bibfield{author}{\bibinfo{person}{Krzysztof~Z Gajos}, \bibinfo{person}{Jacob~O
  Wobbrock}, {and} \bibinfo{person}{Daniel~S Weld}.}
  \bibinfo{year}{2008}\natexlab{}.
\newblock \showarticletitle{Improving the performance of motor-impaired users
  with automatically-generated, ability-based interfaces}. In
  \bibinfo{booktitle}{\emph{Proceedings of the SIGCHI conference on Human
  Factors in Computing Systems}}. \bibinfo{pages}{1257--1266}.
\newblock


\bibitem[Gray and Boehm-Davis(2000)]%
        {gray2000milliseconds}
\bibfield{author}{\bibinfo{person}{Wayne~D Gray} {and}
  \bibinfo{person}{Deborah~A Boehm-Davis}.} \bibinfo{year}{2000}\natexlab{}.
\newblock \showarticletitle{Milliseconds matter: An introduction to
  microstrategies and to their use in describing and predicting interactive
  behavior.}
\newblock \bibinfo{journal}{\emph{Journal of experimental psychology: applied}}
  \bibinfo{volume}{6}, \bibinfo{number}{4} (\bibinfo{year}{2000}),
  \bibinfo{pages}{322}.
\newblock


\bibitem[Holz and Baudisch(2010)]%
        {holz2010generalized}
\bibfield{author}{\bibinfo{person}{Christian Holz} {and}
  \bibinfo{person}{Patrick Baudisch}.} \bibinfo{year}{2010}\natexlab{}.
\newblock \showarticletitle{The generalized perceived input point model and how
  to double touch accuracy by extracting fingerprints}. In
  \bibinfo{booktitle}{\emph{Proceedings of the SIGCHI Conference on Human
  Factors in Computing Systems}}. \bibinfo{pages}{581--590}.
\newblock


\bibitem[Holz and Baudisch(2011)]%
        {holz2011understanding}
\bibfield{author}{\bibinfo{person}{Christian Holz} {and}
  \bibinfo{person}{Patrick Baudisch}.} \bibinfo{year}{2011}\natexlab{}.
\newblock \showarticletitle{Understanding touch}. In
  \bibinfo{booktitle}{\emph{Proceedings of the SIGCHI Conference on Human
  Factors in Computing Systems}}. \bibinfo{pages}{2501--2510}.
\newblock


\bibitem[Jiang et~al\mbox{.}(2019)]%
        {jiang2019orc}
\bibfield{author}{\bibinfo{person}{Yue Jiang}, \bibinfo{person}{Ruofei Du},
  \bibinfo{person}{Christof Lutteroth}, {and} \bibinfo{person}{Wolfgang
  Stuerzlinger}.} \bibinfo{year}{2019}\natexlab{}.
\newblock \showarticletitle{ORC layout: Adaptive GUI layout with
  OR-constraints}. In \bibinfo{booktitle}{\emph{Proceedings of the 2019 CHI
  Conference on Human Factors in Computing Systems}}. \bibinfo{pages}{1--12}.
\newblock


\bibitem[Jiang et~al\mbox{.}(2021)]%
        {jiang2021reverseorc}
\bibfield{author}{\bibinfo{person}{Yue Jiang}, \bibinfo{person}{Wolfgang
  Stuerzlinger}, {and} \bibinfo{person}{Christof Lutteroth}.}
  \bibinfo{year}{2021}\natexlab{}.
\newblock \showarticletitle{ReverseORC: Reverse Engineering of Resizable User
  Interface Layouts with OR-Constraints}. In
  \bibinfo{booktitle}{\emph{Proceedings of the 2021 CHI Conference on Human
  Factors in Computing Systems}}. \bibinfo{pages}{1--18}.
\newblock


\bibitem[Kane et~al\mbox{.}(2008)]%
        {kane2008getting}
\bibfield{author}{\bibinfo{person}{Shaun~K Kane}, \bibinfo{person}{Jacob~O
  Wobbrock}, {and} \bibinfo{person}{Ian~E Smith}.}
  \bibinfo{year}{2008}\natexlab{}.
\newblock \showarticletitle{Getting off the treadmill: evaluating walking user
  interfaces for mobile devices in public spaces}. In
  \bibinfo{booktitle}{\emph{Proceedings of the 10th international conference on
  Human computer interaction with mobile devices and services}}.
  \bibinfo{pages}{109--118}.
\newblock


\bibitem[Kingma and Ba(2014)]%
        {kingma2014adam}
\bibfield{author}{\bibinfo{person}{Diederik~P Kingma} {and}
  \bibinfo{person}{Jimmy Ba}.} \bibinfo{year}{2014}\natexlab{}.
\newblock \showarticletitle{Adam: A method for stochastic optimization}.
\newblock \bibinfo{journal}{\emph{arXiv preprint arXiv:1412.6980}}
  (\bibinfo{year}{2014}).
\newblock


\bibitem[Krosnick et~al\mbox{.}(2018)]%
        {krosnick2018expresso}
\bibfield{author}{\bibinfo{person}{Rebecca Krosnick}, \bibinfo{person}{Sang~Won
  Lee}, \bibinfo{person}{Walter~S Laseck}, {and} \bibinfo{person}{Steve Onev}.}
  \bibinfo{year}{2018}\natexlab{}.
\newblock \showarticletitle{Expresso: Building responsive interfaces with
  keyframes}. In \bibinfo{booktitle}{\emph{2018 IEEE Symposium on Visual
  Languages and Human-Centric Computing (VL/HCC)}}. IEEE,
  \bibinfo{pages}{39--47}.
\newblock


\bibitem[Lakens(2013)]%
        {lakens2013calculating}
\bibfield{author}{\bibinfo{person}{Dani{\"e}l Lakens}.}
  \bibinfo{year}{2013}\natexlab{}.
\newblock \showarticletitle{Calculating and reporting effect sizes to
  facilitate cumulative science: a practical primer for t-tests and ANOVAs}.
\newblock \bibinfo{journal}{\emph{Frontiers in psychology}}
  \bibinfo{volume}{4} (\bibinfo{year}{2013}), \bibinfo{pages}{863}.
\newblock


\bibitem[Le et~al\mbox{.}(2018)]%
        {le2018fingers}
\bibfield{author}{\bibinfo{person}{Huy~Viet Le}, \bibinfo{person}{Sven Mayer},
  \bibinfo{person}{Patrick Bader}, {and} \bibinfo{person}{Niels Henze}.}
  \bibinfo{year}{2018}\natexlab{}.
\newblock \showarticletitle{Fingers' Range and Comfortable Area for One-Handed
  Smartphone Interaction Beyond the Touchscreen}. In
  \bibinfo{booktitle}{\emph{Proceedings of the 2018 CHI Conference on Human
  Factors in Computing Systems}}. \bibinfo{pages}{1--12}.
\newblock


\bibitem[Le et~al\mbox{.}(2019)]%
        {le2019investigating}
\bibfield{author}{\bibinfo{person}{Huy~Viet Le}, \bibinfo{person}{Sven Mayer},
  \bibinfo{person}{Benedict Steuerlein}, {and} \bibinfo{person}{Niels Henze}.}
  \bibinfo{year}{2019}\natexlab{}.
\newblock \showarticletitle{Investigating Unintended Inputs for One-Handed
  Touch Interaction Beyond the Touchscreen}. In
  \bibinfo{booktitle}{\emph{Proceedings of the 21st International Conference on
  Human-Computer Interaction with Mobile Devices and Services}}.
  \bibinfo{pages}{1--14}.
\newblock


\bibitem[Li et~al\mbox{.}(2018)]%
        {li2018predicting}
\bibfield{author}{\bibinfo{person}{Yang Li}, \bibinfo{person}{Samy Bengio},
  {and} \bibinfo{person}{Gilles Bailly}.} \bibinfo{year}{2018}\natexlab{}.
\newblock \showarticletitle{Predicting human performance in vertical menu
  selection using deep learning}. In \bibinfo{booktitle}{\emph{Proceedings of
  the 2018 CHI Conference on Human Factors in Computing Systems}}.
  \bibinfo{pages}{1--7}.
\newblock


\bibitem[Lutteroth(2008)]%
        {lutteroth2008automated}
\bibfield{author}{\bibinfo{person}{Christof Lutteroth}.}
  \bibinfo{year}{2008}\natexlab{}.
\newblock \showarticletitle{Automated reverse engineering of hard-coded GUI
  layouts}. In \bibinfo{booktitle}{\emph{Proceedings of the ninth conference on
  Australasian user interface-Volume 76}}. \bibinfo{pages}{65--73}.
\newblock


\bibitem[MacKenzie(1992)]%
        {mackenzie1992fitts}
\bibfield{author}{\bibinfo{person}{I~Scott MacKenzie}.}
  \bibinfo{year}{1992}\natexlab{}.
\newblock \showarticletitle{Fitts' law as a research and design tool in
  human-computer interaction}.
\newblock \bibinfo{journal}{\emph{Human-computer interaction}}
  \bibinfo{volume}{7}, \bibinfo{number}{1} (\bibinfo{year}{1992}),
  \bibinfo{pages}{91--139}.
\newblock


\bibitem[Mankoff et~al\mbox{.}(2003)]%
        {mankoff2003heuristic}
\bibfield{author}{\bibinfo{person}{Jennifer Mankoff}, \bibinfo{person}{Anind~K
  Dey}, \bibinfo{person}{Gary Hsieh}, \bibinfo{person}{Julie Kientz},
  \bibinfo{person}{Scott Lederer}, {and} \bibinfo{person}{Morgan Ames}.}
  \bibinfo{year}{2003}\natexlab{}.
\newblock \showarticletitle{Heuristic evaluation of ambient displays}. In
  \bibinfo{booktitle}{\emph{Proceedings of the SIGCHI conference on Human
  factors in computing systems}}. \bibinfo{pages}{169--176}.
\newblock


\bibitem[Mariakakis et~al\mbox{.}(2018)]%
        {mariakakis2018drunk}
\bibfield{author}{\bibinfo{person}{Alex Mariakakis}, \bibinfo{person}{Sayna
  Parsi}, \bibinfo{person}{Shwetak~N Patel}, {and} \bibinfo{person}{Jacob~O
  Wobbrock}.} \bibinfo{year}{2018}\natexlab{}.
\newblock \showarticletitle{Drunk user interfaces: Determining blood alcohol
  level through everyday smartphone tasks}. In
  \bibinfo{booktitle}{\emph{Proceedings of the 2018 CHI conference on human
  factors in computing systems}}. \bibinfo{pages}{1--13}.
\newblock


\bibitem[Mayer et~al\mbox{.}(2019)]%
        {mayer2019finding}
\bibfield{author}{\bibinfo{person}{Sven Mayer}, \bibinfo{person}{Huy~Viet Le},
  \bibinfo{person}{Markus Funk}, {and} \bibinfo{person}{Niels Henze}.}
  \bibinfo{year}{2019}\natexlab{}.
\newblock \showarticletitle{Finding the Sweet Spot: Analyzing Unrestricted
  Touchscreen Interaction In-the-Wild}. In
  \bibinfo{booktitle}{\emph{Proceedings of the 2019 ACM International
  Conference on Interactive Surfaces and Spaces}}. \bibinfo{pages}{171--179}.
\newblock


\bibitem[Mott and Wobbrock(2019)]%
        {mott2019cluster}
\bibfield{author}{\bibinfo{person}{Martez~E Mott} {and}
  \bibinfo{person}{Jacob~O Wobbrock}.} \bibinfo{year}{2019}\natexlab{}.
\newblock \showarticletitle{Cluster Touch: Improving touch accuracy on
  smartphones for people with motor and situational impairments}. In
  \bibinfo{booktitle}{\emph{Proceedings of the 2019 CHI Conference on Human
  Factors in Computing Systems}}. \bibinfo{pages}{1--14}.
\newblock


\bibitem[Negulescu and McGrenere(2015)]%
        {negulescu2015grip}
\bibfield{author}{\bibinfo{person}{Matei Negulescu} {and}
  \bibinfo{person}{Joanna McGrenere}.} \bibinfo{year}{2015}\natexlab{}.
\newblock \showarticletitle{Grip change as an information side channel for
  mobile touch interaction}. In \bibinfo{booktitle}{\emph{Proceedings of the
  33rd Annual ACM Conference on Human Factors in Computing Systems}}.
  \bibinfo{pages}{1519--1522}.
\newblock


\bibitem[Nichols et~al\mbox{.}(2008)]%
        {nichols2008highlight}
\bibfield{author}{\bibinfo{person}{Jeffrey Nichols}, \bibinfo{person}{Zhigang
  Hua}, {and} \bibinfo{person}{John Barton}.} \bibinfo{year}{2008}\natexlab{}.
\newblock \showarticletitle{Highlight: a system for creating and deploying
  mobile web applications}. In \bibinfo{booktitle}{\emph{Proceedings of the
  21st annual ACM symposium on User interface software and technology}}.
  \bibinfo{pages}{249--258}.
\newblock


\bibitem[Nielsen(1994)]%
        {10.5555/189200.189209}
\bibfield{author}{\bibinfo{person}{Jakob Nielsen}.}
  \bibinfo{year}{1994}\natexlab{}.
\newblock \bibinfo{booktitle}{\emph{Heuristic Evaluation}}.
\newblock \bibinfo{publisher}{John Wiley \& Sons, Inc.},
  \bibinfo{address}{USA}, \bibinfo{pages}{25–62}.
\newblock
\showISBNx{0471018775}


\bibitem[O'Donovan et~al\mbox{.}(2015)]%
        {o2015designscape}
\bibfield{author}{\bibinfo{person}{Peter O'Donovan}, \bibinfo{person}{Aseem
  Agarwala}, {and} \bibinfo{person}{Aaron Hertzmann}.}
  \bibinfo{year}{2015}\natexlab{}.
\newblock \showarticletitle{Designscape: Design with interactive layout
  suggestions}. In \bibinfo{booktitle}{\emph{Proceedings of the 33rd annual ACM
  conference on human factors in computing systems}}.
  \bibinfo{pages}{1221--1224}.
\newblock


\bibitem[O’Donovan et~al\mbox{.}(2014)]%
        {o2014learning}
\bibfield{author}{\bibinfo{person}{Peter O’Donovan}, \bibinfo{person}{Aseem
  Agarwala}, {and} \bibinfo{person}{Aaron Hertzmann}.}
  \bibinfo{year}{2014}\natexlab{}.
\newblock \showarticletitle{Learning layouts for single-pagegraphic designs}.
\newblock \bibinfo{journal}{\emph{IEEE transactions on visualization and
  computer graphics}} \bibinfo{volume}{20}, \bibinfo{number}{8}
  (\bibinfo{year}{2014}), \bibinfo{pages}{1200--1213}.
\newblock


\bibitem[Platt and Barr(1988)]%
        {platt1988constrained}
\bibfield{author}{\bibinfo{person}{John~C Platt} {and} \bibinfo{person}{Alan~H
  Barr}.} \bibinfo{year}{1988}\natexlab{}.
\newblock \showarticletitle{Constrained differential optimization for neural
  networks}.
\newblock  (\bibinfo{year}{1988}).
\newblock


\bibitem[Sarcar et~al\mbox{.}(2018)]%
        {sarcar2018ability}
\bibfield{author}{\bibinfo{person}{Sayan Sarcar}, \bibinfo{person}{Jussi~PP
  Jokinen}, \bibinfo{person}{Antti Oulasvirta}, \bibinfo{person}{Zhenxin Wang},
  \bibinfo{person}{Chaklam Silpasuwanchai}, {and} \bibinfo{person}{Xiangshi
  Ren}.} \bibinfo{year}{2018}\natexlab{}.
\newblock \showarticletitle{Ability-based optimization of touchscreen
  interactions}.
\newblock \bibinfo{journal}{\emph{IEEE Pervasive Computing}}
  \bibinfo{volume}{17}, \bibinfo{number}{1} (\bibinfo{year}{2018}),
  \bibinfo{pages}{15--26}.
\newblock


\bibitem[Schwerdtfeger(1991)]%
        {outspoken}
\bibfield{author}{\bibinfo{person}{Richard~S. Schwerdtfeger}.}
  \bibinfo{year}{1991}\natexlab{}.
\newblock \bibinfo{title}{Making the GUI Talk}.
\newblock
\newblock
\newblock
\shownote{ftp://service.boulder.ibm.com/sns/sr-os2/sr2doc/guitalk.txt}.


\bibitem[Shiripour et~al\mbox{.}(2021)]%
        {shiripour2021grid}
\bibfield{author}{\bibinfo{person}{Morteza Shiripour},
  \bibinfo{person}{Niraj~Ramesh Dayama}, {and} \bibinfo{person}{Antti
  Oulasvirta}.} \bibinfo{year}{2021}\natexlab{}.
\newblock \showarticletitle{Grid-based Genetic Operators for Graphical Layout
  Generation}.
\newblock \bibinfo{journal}{\emph{Proceedings of the ACM on Human-Computer
  Interaction}} \bibinfo{volume}{5}, \bibinfo{number}{EICS}
  (\bibinfo{year}{2021}), \bibinfo{pages}{1--30}.
\newblock


\bibitem[Smith(2007)]%
        {smith2007overview}
\bibfield{author}{\bibinfo{person}{Ray Smith}.}
  \bibinfo{year}{2007}\natexlab{}.
\newblock \showarticletitle{An overview of the Tesseract OCR engine}. In
  \bibinfo{booktitle}{\emph{Ninth international conference on document analysis
  and recognition (ICDAR 2007)}}, Vol.~\bibinfo{volume}{2}. IEEE,
  \bibinfo{pages}{629--633}.
\newblock


\bibitem[Stuerzlinger et~al\mbox{.}(2006)]%
        {stuerzlinger2006user}
\bibfield{author}{\bibinfo{person}{Wolfgang Stuerzlinger},
  \bibinfo{person}{Olivier Chapuis}, \bibinfo{person}{Dusty Phillips}, {and}
  \bibinfo{person}{Nicolas Roussel}.} \bibinfo{year}{2006}\natexlab{}.
\newblock \showarticletitle{User interface fa{\c{c}}ades: towards fully
  adaptable user interfaces}. In \bibinfo{booktitle}{\emph{Proceedings of the
  19th annual ACM symposium on User interface software and technology}}.
  \bibinfo{pages}{309--318}.
\newblock


\bibitem[Swearngin et~al\mbox{.}(2017)]%
        {swearngin2017genie}
\bibfield{author}{\bibinfo{person}{Amanda Swearngin}, \bibinfo{person}{Amy~J
  Ko}, {and} \bibinfo{person}{James Fogarty}.} \bibinfo{year}{2017}\natexlab{}.
\newblock \showarticletitle{Genie: Input Retargeting on the Web through Command
  Reverse Engineering}. In \bibinfo{booktitle}{\emph{Proceedings of the 2017
  CHI Conference on Human Factors in Computing Systems}}.
  \bibinfo{pages}{4703--4714}.
\newblock


\bibitem[Swearngin et~al\mbox{.}(2020)]%
        {swearngin2020scout}
\bibfield{author}{\bibinfo{person}{Amanda Swearngin},
  \bibinfo{person}{Chenglong Wang}, \bibinfo{person}{Alannah Oleson},
  \bibinfo{person}{James Fogarty}, {and} \bibinfo{person}{Amy~J Ko}.}
  \bibinfo{year}{2020}\natexlab{}.
\newblock \showarticletitle{Scout: Rapid Exploration of Interface Layout
  Alternatives through High-Level Design Constraints}. In
  \bibinfo{booktitle}{\emph{Proceedings of the 2020 CHI Conference on Human
  Factors in Computing Systems}}. \bibinfo{pages}{1--13}.
\newblock


\bibitem[Telea(2004)]%
        {telea2004image}
\bibfield{author}{\bibinfo{person}{Alexandru Telea}.}
  \bibinfo{year}{2004}\natexlab{}.
\newblock \showarticletitle{An image inpainting technique based on the fast
  marching method}.
\newblock \bibinfo{journal}{\emph{Journal of graphics tools}}
  \bibinfo{volume}{9}, \bibinfo{number}{1} (\bibinfo{year}{2004}),
  \bibinfo{pages}{23--34}.
\newblock


\bibitem[Todi et~al\mbox{.}(2016)]%
        {todi2016sketchplore}
\bibfield{author}{\bibinfo{person}{Kashyap Todi}, \bibinfo{person}{Daryl Weir},
  {and} \bibinfo{person}{Antti Oulasvirta}.} \bibinfo{year}{2016}\natexlab{}.
\newblock \showarticletitle{Sketchplore: Sketch and explore with a layout
  optimiser}. In \bibinfo{booktitle}{\emph{Proceedings of the 2016 ACM
  Conference on Designing Interactive Systems}}. \bibinfo{pages}{543--555}.
\newblock


\bibitem[Vogel and Baudisch(2007)]%
        {vogel2007shift}
\bibfield{author}{\bibinfo{person}{Daniel Vogel} {and} \bibinfo{person}{Patrick
  Baudisch}.} \bibinfo{year}{2007}\natexlab{}.
\newblock \showarticletitle{Shift: a technique for operating pen-based
  interfaces using touch}. In \bibinfo{booktitle}{\emph{Proceedings of the
  SIGCHI conference on Human factors in computing systems}}.
  \bibinfo{pages}{657--666}.
\newblock


\bibitem[Wobbrock et~al\mbox{.}(2011)]%
        {wobbrock2011ability}
\bibfield{author}{\bibinfo{person}{Jacob~O Wobbrock}, \bibinfo{person}{Shaun~K
  Kane}, \bibinfo{person}{Krzysztof~Z Gajos}, \bibinfo{person}{Susumu Harada},
  {and} \bibinfo{person}{Jon Froehlich}.} \bibinfo{year}{2011}\natexlab{}.
\newblock \showarticletitle{Ability-based design: Concept, principles and
  examples}.
\newblock \bibinfo{journal}{\emph{ACM Transactions on Accessible Computing
  (TACCESS)}} \bibinfo{volume}{3}, \bibinfo{number}{3} (\bibinfo{year}{2011}),
  \bibinfo{pages}{1--27}.
\newblock


\bibitem[Wu et~al\mbox{.}(2021)]%
        {wu2021screen}
\bibfield{author}{\bibinfo{person}{Jason Wu}, \bibinfo{person}{Xiaoyi Zhang},
  \bibinfo{person}{Jeffrey Nichols}, {and} \bibinfo{person}{Jeffrey~P.
  Bigham}.} \bibinfo{year}{2021}\natexlab{}.
\newblock \showarticletitle{Screen Parsing: Towards Reverse Engineering of UI
  Models from Screenshots}. In \bibinfo{booktitle}{\emph{Proceedings of the
  2021 ACM Symposium on User Interface Software and Technology}}.
  \bibinfo{pages}{1--14}.
\newblock


\bibitem[Yeh et~al\mbox{.}(2009)]%
        {sikuli}
\bibfield{author}{\bibinfo{person}{Tom Yeh}, \bibinfo{person}{Tsung-Hsiang
  Chang}, {and} \bibinfo{person}{Robert~C. Miller}.}
  \bibinfo{year}{2009}\natexlab{}.
\newblock \showarticletitle{Sikuli: Using GUI Screenshots for Search and
  Automation}. In \bibinfo{booktitle}{\emph{Proceedings of the 22nd Annual ACM
  Symposium on User Interface Software and Technology}} (Victoria, BC, Canada)
  \emph{(\bibinfo{series}{UIST ’09})}. \bibinfo{publisher}{Association for
  Computing Machinery}, \bibinfo{address}{New York, NY, USA},
  \bibinfo{pages}{183–192}.
\newblock
\showISBNx{9781605587455}
\urldef\tempurl%
\url{https://doi.org/10.1145/1622176.1622213}
\showDOI{\tempurl}


\bibitem[Zeidler et~al\mbox{.}(2013)]%
        {zeidler2013auckland}
\bibfield{author}{\bibinfo{person}{Clemens Zeidler}, \bibinfo{person}{Christof
  Lutteroth}, \bibinfo{person}{Wolfgang Sturzlinger}, {and}
  \bibinfo{person}{Gerald Weber}.} \bibinfo{year}{2013}\natexlab{}.
\newblock \showarticletitle{The auckland layout editor: an improved GUI layout
  specification process}. In \bibinfo{booktitle}{\emph{Proceedings of the 26th
  annual ACM symposium on User interface software and technology}}.
  \bibinfo{pages}{343--352}.
\newblock


\bibitem[Zeidler et~al\mbox{.}(2017)]%
        {zeidler2017automatic}
\bibfield{author}{\bibinfo{person}{Clemens Zeidler}, \bibinfo{person}{Gerald
  Weber}, \bibinfo{person}{Wolfgang Stuerzlinger}, {and}
  \bibinfo{person}{Christof Lutteroth}.} \bibinfo{year}{2017}\natexlab{}.
\newblock \showarticletitle{Automatic generation of user interface layouts for
  alternative screen orientations}. In \bibinfo{booktitle}{\emph{IFIP
  Conference on Human-Computer Interaction}}. Springer,
  \bibinfo{pages}{13--35}.
\newblock


\bibitem[Zhang et~al\mbox{.}(2021)]%
        {zhang2021screen}
\bibfield{author}{\bibinfo{person}{Xiaoyi Zhang}, \bibinfo{person}{Lilian de
  Greef}, \bibinfo{person}{Amanda Swearngin}, \bibinfo{person}{Samuel White},
  \bibinfo{person}{Kyle Murray}, \bibinfo{person}{Lisa Yu}, \bibinfo{person}{Qi
  Shan}, \bibinfo{person}{Jeffrey Nichols}, \bibinfo{person}{Jason Wu},
  \bibinfo{person}{Chris Fleizach}, {et~al\mbox{.}}}
  \bibinfo{year}{2021}\natexlab{}.
\newblock \showarticletitle{Screen Recognition: Creating Accessibility Metadata
  for Mobile Applications from Pixels}. In
  \bibinfo{booktitle}{\emph{Proceedings of the 2021 CHI Conference on Human
  Factors in Computing Systems}}. \bibinfo{pages}{1--15}.
\newblock


\bibitem[Zhang et~al\mbox{.}(2017)]%
        {zhang2017interaction}
\bibfield{author}{\bibinfo{person}{Xiaoyi Zhang}, \bibinfo{person}{Anne~Spencer
  Ross}, \bibinfo{person}{Anat Caspi}, \bibinfo{person}{James Fogarty}, {and}
  \bibinfo{person}{Jacob~O Wobbrock}.} \bibinfo{year}{2017}\natexlab{}.
\newblock \showarticletitle{Interaction proxies for runtime repair and
  enhancement of mobile application accessibility}. In
  \bibinfo{booktitle}{\emph{Proceedings of the 2017 CHI Conference on Human
  Factors in Computing Systems}}. \bibinfo{pages}{6024--6037}.
\newblock


\end{thebibliography}
